\documentclass[11pt]{article}
\pdfoutput=1

\usepackage{jheppub}
\usepackage{amsmath, amsfonts, amssymb}
\usepackage{graphicx}
\usepackage{psfrag}
\usepackage[usenames,dvipsnames,svgnames,table]{xcolor}
\usepackage{enumerate}
\usepackage{soul}
\usepackage{hyperref}
\usepackage{subfig}
\usepackage{comment}
\usepackage[T1]{fontenc}

\newcommand{\z}{\zeta}

\newcommand{\A}{A}
\newcommand{\bend}{\text{turn}}
\newcommand{\turn}{\text{turn}}
\newcommand{\<}{\langle}
\renewcommand{\>}{\rangle}
\newcommand{\be}{\begin{equation}}
\newcommand{\ee}{\end{equation}}
\newcommand{\lwick}{:\!}
\newcommand{\rwick}{\!:}

\def\be{\begin{equation}}
\def\ee{\end{equation}}
\def\beq{\begin{equation}}
\def\eeq{\end{equation}}

\newcommand{\al}{\alpha}
\newcommand{\bt}{\beta}

\renewcommand{\(}{\left(}
\renewcommand{\)}{\right)}

\newcommand{\e}{\text{e}}

\renewcommand{\O}{\mathcal{O}}

\renewcommand{\L}{\mathcal{L}}

\newcommand{\I}{i}
\newcommand{\D}{\mathrm{d}}
\newcommand{\C}{\mathbb{C}}
\newcommand{\nn}{\nonumber}

\DeclareMathOperator{\arccosh}{arccosh}

\DeclareMathOperator{\arctanh}{arctanh}

\newcommand{\figref}[1]{Fig. \ref{#1}}
\newcommand{\secref}[1]{Sec. \ref{#1}}

\def\simleq{\; \raise0.3ex\hbox{$<$\kern-0.75em
      \raise-1.1ex\hbox{$\sim$}}\; }
   \def\simgeq{\; \raise0.3ex\hbox{$>$\kern-0.75em
      \raise-1.1ex\hbox{$\sim$}}\; }

\title{Cosmological singularities\\encoded in IR boundary correlations}
\author{Adam Bzowski,}
\author{Thomas Hertog,}
\author{and Marjorie Schillo}

\emailAdd{adam.bzowski@fys.kuleuven.be}
\emailAdd{thomas.hertog@fys.kuleuven.be}
\emailAdd{marjorie@itf.fys.kuleuven.be}

\affiliation{\it Institute for Theoretical Physics, KU Leuven, 3001 Leuven, Belgium }

\abstract{We study the dynamics near big crunch singularities produced in asymptotic $AdS$ cosmologies using gauge/gravity duality. The dual description consists of a constant mass deformation of ABJM theory on de Sitter space and is well-defined and stable for small deformations. We identify the critical deformation where the theory becomes unstable at weak and at strong coupling. Using spacelike geodesics anchored on the boundary we compute two-point correlators of ABJM operators of large dimensions. Near the critical deformation a second saddle point contribution enters, in which the spacelike geodesics probe the high curvature region near the singularity. Its contribution strongly enhances  the long-distance correlations. This has a natural interpretation in the weakly coupled boundary theory where the critical point corresponds to a massless limit.}

\begin{document}

\maketitle

\section{Introduction}
A longstanding goal of quantum gravity is to describe the physics near singularities such as the big bang or inside black holes. 
Gauge/gravity duality provides a powerful tool to apply to this problem since it allows us to describe singularities in toy model cosmologies that are asymptotically anti-de Sitter ($AdS$) in terms of a dual quantum field theory living on the conformal boundary. 

The first examples of such `$AdS$ cosmologies' were constructed in \cite{Hertog:2004rz,Hertog:2005hu}. These are solutions of $\mathcal{N}=8$, $D=4$ supergravity involving only gravity and a single scalar field with $m^2 = -2/R_{AdS}^2$.  In these scenarios smooth, spherically symmetric, asymptotically $AdS$ initial data evolves into (or from) a singularity which extends all the way  to infinity. The Lorentzian evolution exhibits an enhanced $O(3,1)$ symmetry characteristic of open Friedman-Lema\^{i}tre-Robertson-Walker (FLRW) cosmologies. The dynamics of the system are governed by the scalar, which  rolls down its (negative) potential, producing a big crunch singularity in finite time. Models of this type were further explored in \cite{Turok:2007ry,Barbon:2011ta,Smolkin:2012er, Kumar:2015jxa} and other models of cosmologies in $AdS$ were studied \emph{e.g.}, in \cite{Craps:2006xq,Das:2006dz,Awad:2008jf,Engelhardt:2014mea,Engelhardt:2015gta}.

For the $AdS$ cosmologies in \cite{Hertog:2004rz} it was shown that if one defines the dual ABJM theory \cite{Aharony:2008ug} on the global $AdS$ boundary the field theory also becomes singular when the bulk singularity reaches the boundary \cite{Hertog:2005hu,Turok:2007ry}. To evolve the field theory beyond this point requires an additional rule that specifies the transition of the field theory state across the singularity. No convincing rule has been put forward, however, and there is some evidence that this is in fact not possible. Specifically, detailed studies of the boundary evolution \cite{Hertog:2005hu,Battarra:2010id,Turok:2007ry}, as well as more general arguments based on holography \cite{Engelhardt:2015gla}, indicate that there can be no domain beyond the big crunch where the spacetime behaves classically again.

Thus, it is appropriate to adopt a boundary viewpoint in which the bulk singularities lie in the infinite future (or past). Such a framework was put forward in \cite{Maldacena:2010un,Harlow:2010my} where the $AdS$ cosmologies of \cite{Hertog:2004rz,Hertog:2005hu} were reinterpreted as being dual to a field theory on de Sitter spacetime. In this setup the bulk singularity corresponds to the dual field theory state in the asymptotic future (or past), leaving the boundary spacetime regular. Furthermore, the field theory is globally well-defined: even though the bulk scalar field turns on a (homogeneous) negative mass deformation in the dual, the conformal coupling to the de Sitter boundary geometry ensures the deformed theory is stable for sufficiently small deformations. 

The dual description on de Sitter therefore provides an appealing setup to explore the quantum dynamics near cosmological singularities. At the same time it indicates there is an instability for large negative deformations. In particular, there is a critical deformation for which the negative mass deformation by a scalar operator, $\O$, exactly cancels its positive conformal coupling, resulting in new massless excitations. For larger negative deformations the dual theory is unstable, just as one would expect for a tachyonic free theory. As we will review, at the critical point in a free theory the expectation value of ${\cal O}$ diverges and its two-point function develops a strong IR tail. 

In this paper we identify the onset of this instability at strong coupling using gauge/gravity duality. Moreover, we find that observables in the dual field theory near criticality can exhibit clear and strong signatures of the bulk singularity. In particular, we show that, similar to the free theory, the two-point function of large dimension operators on the boundary at strong coupling is  enhanced in the IR as one approaches the critical point, and we trace this feature to the presence of the bulk singularity.

The starting point for our analysis is the construction of an effective potential for the expectation value $\<\O\>$ at strong coupling.  This effective potential is obtained using the asymptotic behavior of the $AdS$ cosmologies together with the bulk boundary conditions. The effective potential shows that there is a critical deformation at which the boundary theory at strong coupling becomes unstable\footnote{The construction of the effective potential is a generalization of the method introduced in \cite{Hertog:2004ns} for duals defined on the global $AdS$ boundary, which used the asymptotic behavior of static scalar soliton solutions. Static solitons are vacua of the theory for certain choices of boundary conditions on global $AdS$. The highly symmetric cosmological backgrounds of \cite{Hertog:2004rz,Hertog:2005hu} are vacua of the theory for the asymptotic de Sitter boundary conditions we consider here. This is also born out by the fact that they can be obtained by analytic continuation of regular Euclidean instantons.}.

To probe and explore the dynamics near the singularity, we compute the two-point function of boundary operators. In the large $N$ limit, the leading contribution to the two-point correlator of an operator, $\mathcal{O}_{\Delta}$, of high conformal dimension, $\Delta$, is specified by the (regulated) length of spacelike bulk geodesics anchored on the boundary\footnote{See \emph{e.g.}, \cite{Fidkowski:2003nf,Festuccia:2005pi} for attempts to probe the singularity inside $AdS$ black holes using geodesics with endpoints on the boundary.} \cite{Balasubramanian:1999zv, Louko:2000tp}.
For small deformations there is a unique bulk background. In this regime all bulk geodesics with endpoints on the boundary stay well away from the high curvature region near the singularity \cite{Engelhardt:2013tra}, and the resulting boundary two-point correlators do not appear to exhibit strong signatures that can be associated with the singularity.

However, larger deformations near the critical point admit a second background. Both backgrounds provide saddle point contributions to the boundary correlators in the geodesic approximation. The second saddle point describes the evolution of initial data consisting of a thin wall bubble with an interior region where the scalar field is large. As a consequence, the singularity develops rapidly in this background. As we will discuss this implies that there are bulk geodesics anchored on the boundary that come close to the singularity. Moreover, their contribution turns out to dominate the large distance behavior of the boundary two-point function, despite the fact that the saddle point is suppressed in the bulk state under consideration,  which is specified by Euclidean initial conditions. In particular, the second saddle point  amplifies correlations in the IR and leads to an IR divergence in the limit of the critical deformation. This limit is probed by spacelike geodesics with endpoints on the boundary that touch the singularity in the bulk. In fact, the IR correlator behaves not unlike the long-distance two-point function of a massless field in de Sitter space, which is precisely what one expects based on the free boundary theory where the critical point corresponds to a massless limit.

\section{Setup} \label{setup}

Our starting point is the low energy limit of M-Theory compactified to $AdS_4 \times S^7$. The massless sector of the theory is $\mathcal{N}=8$ gauged supergravity in four dimensions, which involves the graviton, 28 $SO(8)$ gauge bosons, and 70 real scalars. We consider a particular consistent truncation that includes only Einstein gravity coupled to a single scalar, $\phi$, with potential \cite{Duff:1999gh},
\be\label{potential}
V(\phi) = -2-  \cosh\(\sqrt{2}\,\phi\) \,,
\ee
where we have chosen the gauge coupling so that the $AdS_4$ solution at $\phi=0$ has radius one. The potential \eqref{potential} has a maximum at $\phi =0$ and is unbounded below, but the scalar is within the Breitenlohner-Freedman bound \cite{Breitenlohner:1982jf} and with appropriate asymptotic boundary conditions the $AdS$ vacuum is non-perturbatively stable \cite{Gibbons:1983aq}.

We  define the dual field theory on three-dimensional de Sitter space. Therefore, it is convenient to work in coordinates in which slices of $AdS$ at constant radius, $\rho$, are three-dimensional de Sitter spaces.  In these coordinates the $AdS$ metric is:
\be \label{dS}
ds^2 = d\rho^2 + \sinh^2(\rho)(-d\tau^2 + \cosh^2(\tau)d\Omega_2^2)\ .
\ee
The scalar field equation of motion on this background implies that  the scalar decays at large radius as
\be
\label{asscalar}
\phi = 2\alpha  e^{-\rho}  + 4\beta e^{-2\rho}  + \dots\,,
\ee
where $\alpha$ and $\beta$ may depend on the boundary coordinates.

In order to specify the theory completely, one must fix boundary conditions. The scalar boundary conditions introduced in \cite{Breitenlohner:1982jf}, for which global $AdS$ is stable, correspond to either $\alpha=0$ or $\beta=0$. These are a particular case of Dirichlet or von Neumann boundary conditions respectively. However, there are other possible boundary conditions for which the theory is well-defined and some or all of the asymptotic $AdS$ symmetries are preserved. Most generally, a choice of boundary conditions amounts to specifying both a relation $\beta(\alpha)$ for the asymptotic scalar profile, and a consistent set of falloff conditions on all metric components for which the conserved charges are well-defined \cite{Henneaux:2004zi, Hertog:2004rz, Hertog:2005hu}. In what follows, we adopt von Neumann scalar boundary conditions where $\beta$ is a nonzero constant together with a set of consistent boundary conditions on the metric.

\subsection{Big bang/big crunch cosmologies in AdS} \label{solsec}
We now review some aspects of the $AdS$ cosmologies introduced in \cite{Hertog:2004rz, Hertog:2005hu}. Following \cite{Maldacena:2010un}, we interpret these cosmologies as solutions with $\beta = {\rm constant}$ boundary conditions in the coordinates \eqref{dS}. In this setup the dual field theory lives on three-dimensional de Sitter space and is globally well-defined.

The $AdS$ cosmologies
are obtained from initial data given by the analytic continuation of regular, $O(4)$-invariant, Euclidean instantons of the form
\begin{equation} \label{e:eu_metric}
ds_E^2 = d\rho^2 + A^2(\rho)(d\theta^2 + \sin^2(\theta) d\Omega_2^2)\,.
\end{equation}
The instantons are fully specified by $A(\rho)$ and $\phi(\rho)$, whose dynamics are governed by the field equations derived from the Euclidean action:
\be
\label{Eucaction}
S_E =  \int d^4x \sqrt{g_E} \(-{1 \over 2} R + {1 \over 2}(\nabla \phi)^2 + V(\phi)\right), \, 
\ee
with $V$ given in \eqref{potential} and units where $8\pi G = 1$.

The Euclidean construction of initial conditions is appealing because it provides some information about the underlying state of the bulk which will be important when we study the singularity with boundary probes.  We note that although the solutions are instantons, and the equations of motion are identical to those for Coleman-de Luccia instantons \cite{Coleman:1980aw}, they should not be thought of as $AdS$ vacuum decays. This is because the boundary conditions of the vacuum, $\beta=\alpha=0$, are incompatible with our boundary conditions $\beta=$ constant.

\begin{figure}[ht]
\begin{center}$
\begin{array}{|c | c| }
\hline
	\includegraphics[angle=0, width=0.48\textwidth]{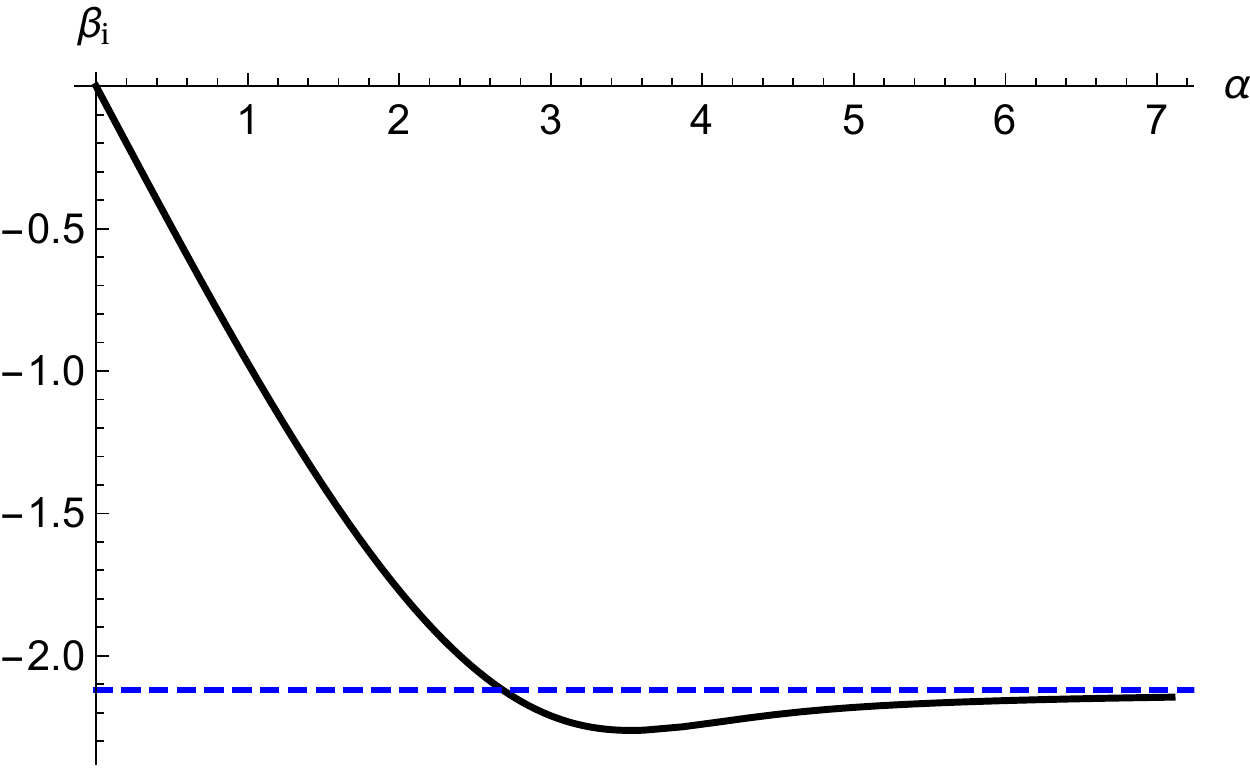} & \includegraphics[angle=0, width=0.48\textwidth]{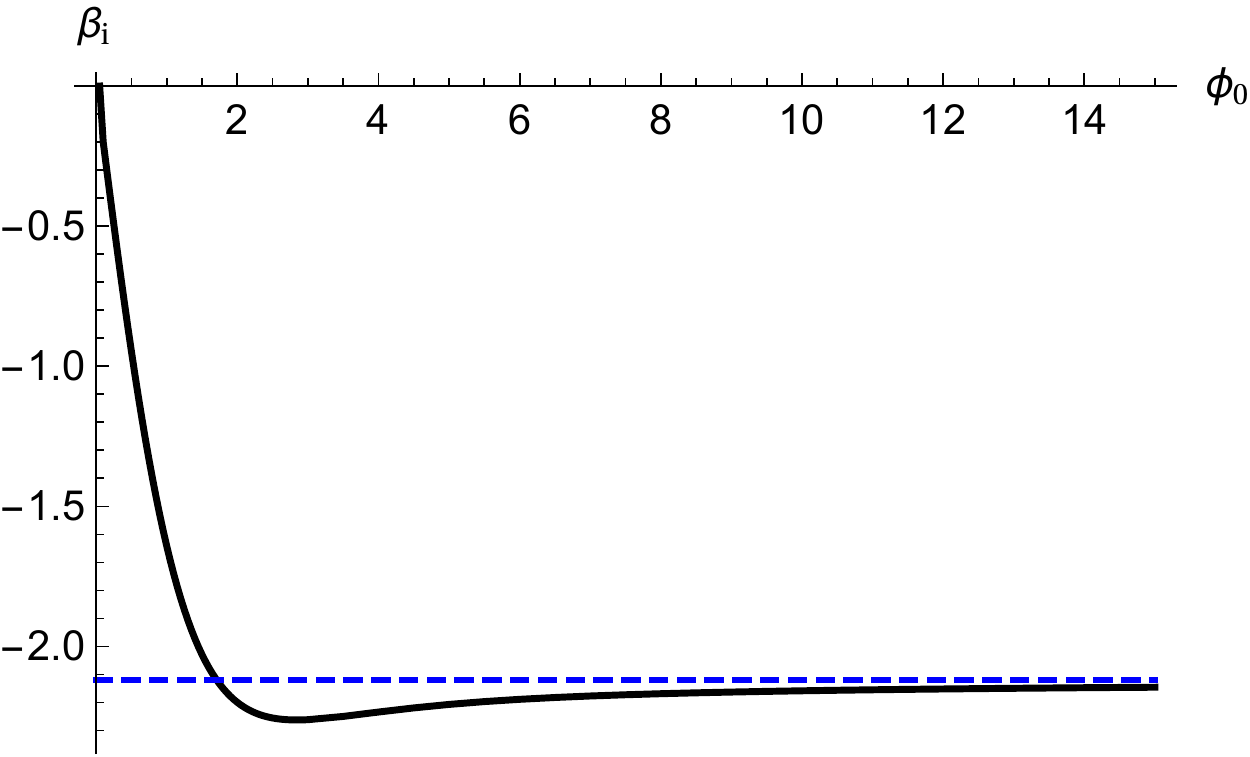} \\
\hline
\end{array}$
\end{center}
\caption{The coefficients that characterize the asymptotic scalar field profile $\beta_{i}(\alpha)$ (left) of the one-parameter family of $O(4)$-invariant, regular instantons, and the relation $\beta_{i}(\phi_0)$ (right), where $\phi_0$ is the value of the scalar at the origin. The fact that $\bt$ tends to a finite value in the large $\al$ limit depends on the detailed form of the scalar potential \eqref{potential} in the consistent truncation we consider.}
\label{betaalpha}
\end{figure}

The field equations determine the asymptotic behavior of the fields\footnote{This definition of $\alpha$ and $\bt$ is consistent with \eqref{asscalar}, where $a_1=1/2$ for empty $AdS$. Note that $a_1 \ne 1/2$ is compatible with an asymptotic $AdS$ structure with radius one.}:
\begin{align}
A(\rho) &= a_{1} \e^\rho + a_{-1}\e^{-\rho} + \mathcal{O}(e^{-2\rho})~, \label{aysa} \\
\phi(\rho) &= {\alpha \over A(\rho)} + {\beta \over A^2(\rho)} + \mathcal{O}(A^{-3}) \nn\\
& = \frac{\alpha}{a_1} e^{-\rho} + \frac{\beta}{a_1^2} e^{-2 \rho} + \mathcal{O}(e^{-3 \rho})~. \label{aysphi}
\end{align}
Regularity at the origin requires $A(0) = 0$, $A'(0) = 1$, and $\phi'(0) =0$. Thus the set of regular instanton solutions  can be labeled by the value of $\phi$ at the origin, $\phi(0)\equiv \phi_0$.  For each $\phi_0$, one can integrate the Einstein equations to find an instanton. Therefore, for each $\phi_0$ one obtains a point in the $(\al,\bt)$ plane and a unique profile for the scale factor. Hence $a_1$ is determined by $\phi_0$ or equivalently $\al$. Repeating for all $\phi_0$ yields a curve $\bt_i(\al)$ where the subscript indicates that this is associated with instanton solutions. We plot this curve\footnote{Since the potential, $V(\phi)$, is even, it suffices to consider positive $\phi_0$ which corresponds to positive $\al$.} in \figref{betaalpha}.

The slice through the instanton at the equator of the $S^3$ in \eqref{e:eu_metric} defines time symmetric initial data for a Lorentzian solution. For any choice of boundary condition $\bt(\al)$, valid initial data are given by the fields on the equatorial slice of the instanton that corresponds to a point where the curve $\bt_i(\al)$ intersects $\bt(\al)$. As can be seen from \figref{betaalpha}, for constant $\beta$ boundary conditions neither existence nor uniqueness of an instanton is guaranteed. There is a minimal value $\beta = \beta_{\rm min}$, below which there are no regular solutions. Furthermore, below some critical value $\beta=\beta_c$ (the dashed line in \figref{betaalpha}) the solution is not unique; boundary conditions $\bt \in (\beta_{\rm min},\beta_c)$ admit two instantons. In this regime Euclidean initial conditions specify a bulk state that is a superposition of two classical backgrounds. For $\beta \rightarrow \beta_c$ the second instanton becomes singular, with divergent $\phi_0$ and $\alpha$.

\begin{figure}[ht]
\begin{center}
	\includegraphics[angle=0, width=0.65\textwidth]{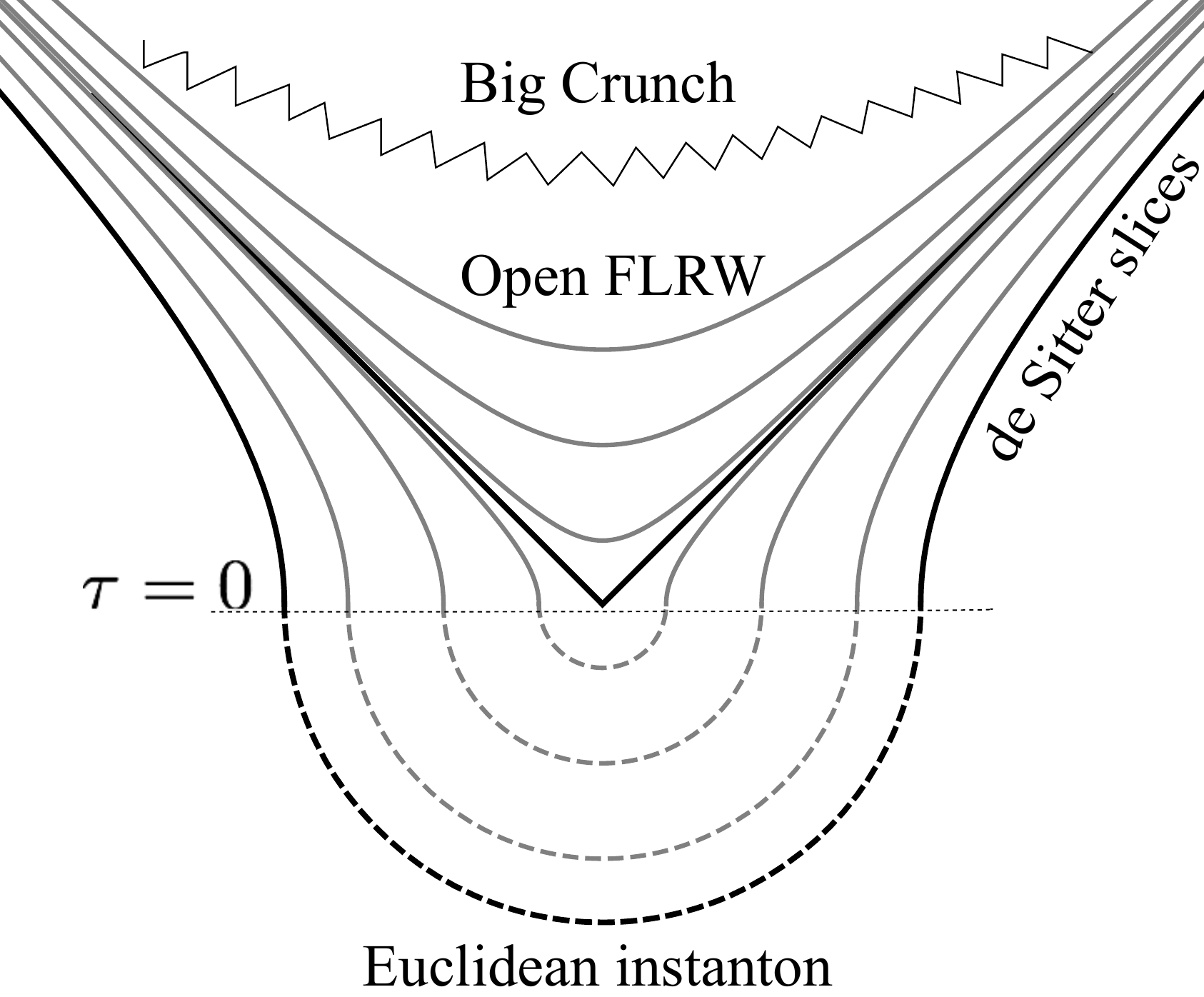}
\end{center}
\caption{Representation of a homogeneous and isotropic, open, asymptotically $AdS$ cosmology with Euclidean initial conditions.  The solution develops a singularity in the interior of the lightcone expanding from the origin. Outside the lightcone the solution is everywhere regular.}
\label{crunchgeo}
\end{figure}

The Lorentzian evolution is obtained by two analytic continuations \cite{Coleman:1980aw,Hertog:2004rz, Hertog:2005hu,Dong:2011gx}: one of which covers the interior of a lightcone emanating form the origin, and one of which covers the exterior of this lightcone. The interior region inherits an $SO(3,1)$ symmetry from the instanton and hence evolves as an open FLRW universe. In the exterior region, spacetime is homogeneous on the radial de Sitter slices discussed above\footnote{For an in-depth discussion of general geometries of this type, see \cite{Dong:2011gx}.} (see \figref{crunchgeo}). We write the Lorentzian metrics inside and outside the lightcone as: 
\begin{align}
ds_{\rm in}^2 &= -dt^2 + a_{\rm in}(t)^2(d\chi^2 + \sinh^2(\chi)d\Omega_2^2)~,\label{interiormetric}\\
ds_{\rm out}^2 &= d\rho^2 + a_{\rm out}(\rho)^2(-d\tau^2 + \cosh^2(\tau)d\Omega_2^2)~. \label{exteriormetric}
\end{align}
The scale factor and the scalar field in the exterior region of the Lorentzian solution do not transform under the analytic continuation from the Euclidean, so $A(\rho) = a_{\rm out}(\rho)$. In particular, the scalar field remains everywhere bounded in this region, and the asymptotic expansions \eqref{aysa} and \eqref{aysphi} remain valid. Hence the Lorentzian evolution given by analytic continuation obeys the constant $\bt$ boundary conditions.

Since both patches are obtained from the analytic continuation of a regular Euclidean geometry, one can use a single complex function to describe both regions. We  define the scale factor, $a$, and the scalar, $\phi$, to be equal to their corresponding values on the inside of the lightcone. The time coordinate $t$ is complex, and its relation to the radial variable is $\rho = i t$. In what follows, we often drop the ``in/out" subscript and use the following notation for the general complex functions:
\be \label{e:aOutToIn}
\begin{split} 
a(t) & = a_{\text{in}}(t) = - i a_{\text{out}}(it)~, \\
\phi(t) & = \phi_{\text{in}}(t) = \phi_{\text{out}}(i t)~.
\end{split}
\ee
The Einstein and Klein-Gordon equations using these conventions can be written:
\begin{align}
{1\over 2} \phi'^2 & = \left({a' \over a}\right)^2-{1\over a^2}-{a'' \over a}~, \label{e:fr1} \\
- {1 \over 2} V(\phi) &= {1\over a^2} - \left({a' \over a}\right)^2 - {a'' \over 2a}~, \label{e:fr2}\\
0 &= \phi'' + \frac{3 a'}{a} \phi' + {d V \over d \phi}~, \label{e:kg}
\end{align}
where prime denotes the derivative with respect to $t$. The Euclidean regularity conditions translate to regularity across the lightcone:
\begin{equation} \label{e:regularity}
a(-t) = -a(t)~, \qquad a'(0) = 1~, \qquad \phi(-t) = \phi(t)~.
\end{equation}

While the scalar field outside the lightcone remains bounded, inside the lightcone $\phi$ rolls down the negative potential. This causes the scale factor $a(t)$ to vanish in finite time $t_c$, producing a big crunch singularity (\emph{cf.}~Appendix \ref{app:crunch}). The lightcone and the singularity within it reach the boundary of $AdS$ in a finite global time \cite{Hertog:2004rz, Hertog:2005hu}. The de Sitter slices however cover only the region outside the lightcone and thus are  everywhere regular. This has important implications for the stability of the dual boundary system as we discuss below.

\begin{figure}[ht]
\begin{center}$
\begin{array}{|c | c| }
\hline
	\includegraphics[angle=0, width=0.48\textwidth]{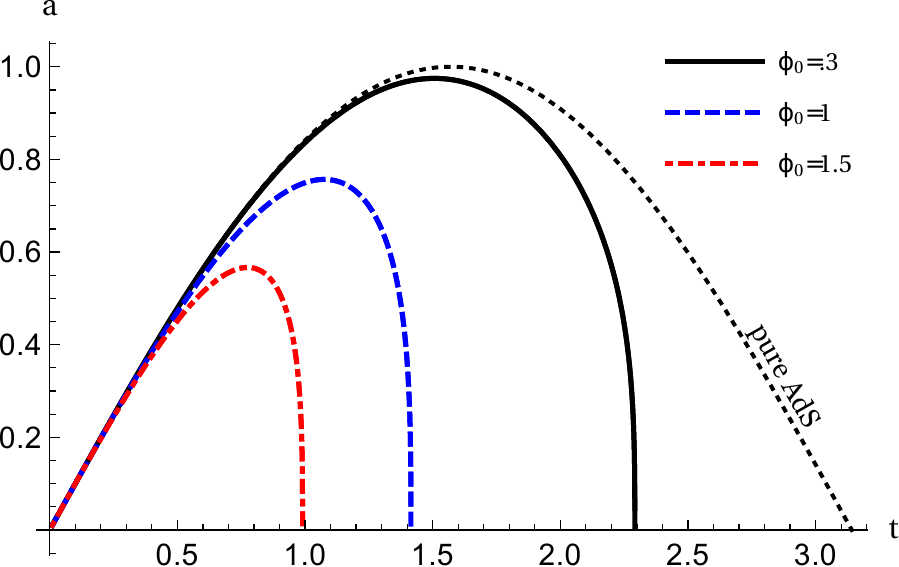} & \includegraphics[angle=0, width=0.48\textwidth]{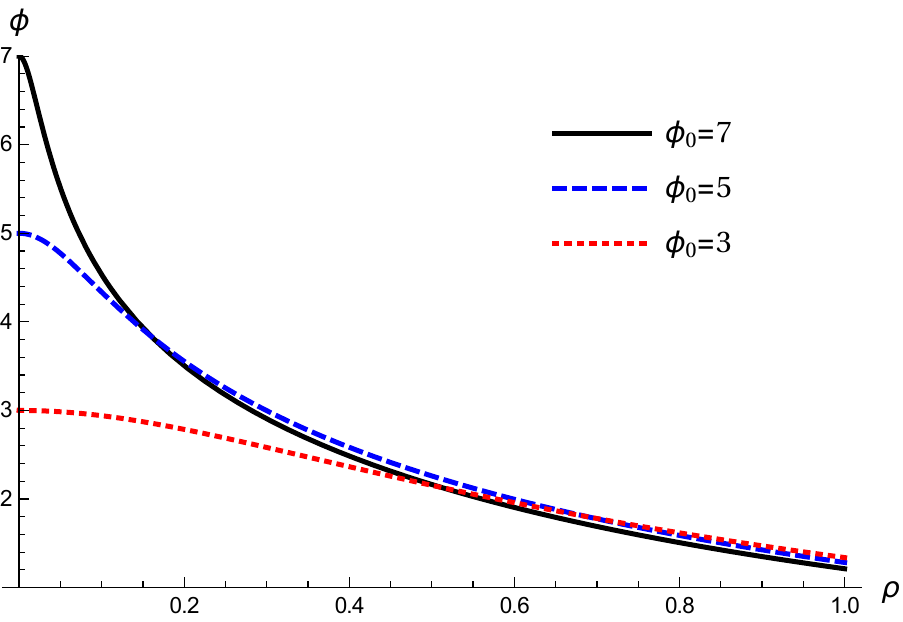} \\
\hline
\end{array}$
\end{center}
\caption{Left panel: Scale factor solutions in the interior for increasing $\phi_0$.  Right panel: Scalar field profiles in the outer region for several large $\phi_0$ showcasing the crossing of solutions.}
\label{geosols}
\end{figure}

In Appendix \ref{app:crunch} (\emph{cf.}~\eqref{e:abound}) we show that the scale factor in the interior region is bounded from above:
\begin{equation} \label{aboundtext}
a(t) <  R\sin\(t/R\) < \sin t \qquad \text{where  } R=\sqrt{{3\over |V(\phi_0)|}}\le 1\,.
\end{equation}
Hence, its maximum, $a_{\rm max} = a(t_{\rm max})$, is bounded by $a_{\rm max}\le R$ with $t_{\max} \le R \pi/2$.  The proper time between the $a=a_{\rm max}$ surface and the singularity is given by $t_c-t_{\max}$. Using steps identical to those leading to \eqref{aboundtext} in Appendix \ref{app:crunch}, it is straightforward to prove that this is also bounded:
\be \label{propDtosing}
t_c-t_{\max} \le {\pi \over 2} a_{\rm max}\,.
\ee
Thus, as we increase $\phi_0$, the proper distance between the surface where the scale factor is maximized and the singularity becomes exponentially small (see also \figref{geosols} (left)). This will be important below when we probe the singularity with geodesics, which do not extend past the surface of maximal scale factor \cite{Engelhardt:2013tra}.  

Finally, we point out a peculiar feature of the instanton solutions, which appears to be a property of this particular truncation of the supergravity theory; namely,  the scalar field profiles, $\phi (\rho)$, for different $\phi_0$ intersect each other. In particular profiles starting at a larger value of $\phi_0$ describe smaller bubbles that are more localized around the origin (\emph{cf.}~\figref{geosols} (right)). Other consistent truncations to a single scalar \cite{Duff:1999gh} do not exhibit this behavior, which appears to be a prerequisite for $\beta_{i}(\alpha)$ to asymptote to a finite value, as seen in \figref{betaalpha} (right). 

\section{Dual field theory} \label{CFTsec}
The bulk system described by the Lorentzian continuation of the action \eqref{Eucaction} has a holographically dual description in terms of a deformation of ABJM theory \cite{Aharony:2008ug} defined on three-dimensional de Sitter space \cite{Maldacena:2010un}. The consistent truncation to a single bulk scalar with $m^2=-2$ corresponds to a single operator in the dual theory of dimension $\Delta$ satisfying the standard relation $m^2 = \Delta(\Delta - d)$.

The scalar boundary conditions in the bulk determine the deformation of the dual field theory. Dirichlet boundary conditions keep the coefficient $\alpha$ in the expansion \eqref{aysphi} fixed and the resulting deformation has $\Delta=2$, while von Neumann boundary conditions keep $\beta$ fixed and deform the boundary theory by an operator with weight $\Delta=1$. More general `mixed' boundary conditions specified by a relation $\beta(\alpha)$ are also allowed \cite{Witten:2001ua,Papadimitriou:2007sj}. The deformation of the dual corresponding to boundary conditions $\beta = \beta(\alpha)$ is given by  \cite{Witten:1998qj, Berkooz:2002ug, Sever:2002fk}
\begin{equation} \label{e:Sdeform}
S_\beta = S_{\text{ABJM}} + \int \D^3 x f(\O_1)~, \qquad\qquad f'(\alpha) = \beta(\alpha)\,,
\end{equation}
where $\al = \< \O_1 \>$. Hence, with our choice of von Neumann boundary conditions $\beta = {\rm constant}$, the boundary theory is simply a massive deformation of ABJM,
\begin{equation} \label{e:deformABJM}
S = S_{\text{ABJM}} + \beta \int \D^3 x \sqrt{-\gamma_{dS}} \O_1\,,
\end{equation}
where $\gamma_{{dS}}$ is the determinant of the metric on the boundary de Sitter space with unit radius.

In the next subsection we calculate the expectation value of the deformation operator, $\<\O_1\>$ via holographic renormalization. We then move on to an analysis of ABJM in the weakly coupled limit in which it reduces to a theory of free scalars on de Sitter space. A free field theory should correspond to a Vasiliev-like, strongly coupled bulk geometry \cite{Fradkin:1986qy, Klebanov:2002ja, Maldacena:2011jn}, so we do not expect to find an exact correspondence with the dynamics in the bulk in the supergravity regime. However, we find that the free theory captures several features of the bulk dynamics in qualitative terms. It therefore provides a framework in which we can interpret and understand some of the holographic signatures of the singularity in the strongly coupled regime that we compute in Section \ref{georesults}, and hence a starting point to explore the quantum dynamics of singularities.

\subsection{Holographic renormalization} \label{sec:HR}
To extract the one-point function of the dual operator $\O_1$, as well as the renormalized value of the on-shell action, holographic renormalization must be carried out \cite{deHaro:2000xn,Bianchi:2001kw,Papadimitriou:2004ap}. To do so it is convenient to work in Euclidean signature, where the line element takes the form \eqref{e:eu_metric}, and the boundary is a three-sphere. The Euclidean  action \eqref{Eucaction}, including the Gibbons-Hawking term, is given by
\begin{equation} \label{e:SEU}
S_E =  \int d^4x \sqrt{g_E} \(-{1 \over 2}R +{1\over 2}(\nabla \phi)^2 + V(\phi)\right) - \int_{\rho = \rho_{\text{cut}}} \D^3 x \sqrt{\gamma^\rho} K\,,
\end{equation}
where $\gamma^\rho_{ij}$ denotes the induced metric on a slice of constant $\rho$, and $K$ is the extrinsic curvature (in the coordinates \eqref{e:eu_metric}, $K = 3 \A'/\A$). In order for the action to be finite we introduce a cut-off, $\rho_{\text{cut}}$. Using the asymptotic expansions \eqref{aysa} and \eqref{aysphi}  for the scale factor  and scalar, one can extract the divergences of the on-shell action,
\begin{equation} \label{e:SEdiv}
S_{\text{div}} = - 2 a_{1}^3 e^{3 \rho_{\text{cut}}} + \frac{a_{1}}{4} e^{\rho_{\text{cut}}} ( \alpha^2 - 6 ) + \text{finite terms}\,.
\end{equation}
A covariant counterterm action removing the divergences reads
\begin{equation} \label{e:Sct}
S_{\text{ct}} = \int_{\rho = \rho_{\text{cut}}} \D^3 x \sqrt{\gamma^{\rho}} \left( 2 + \frac{1}{2} R[\gamma^{\rho}] + \frac{1}{2} \phi^2 \right)\,,
\end{equation}
in agreement with \cite{Papadimitriou:2004ap}.

In the standard quantization scheme, the source for the dual operator of dimension $\Delta = 2$ is identified with the leading term of the expansion of the scalar field, in this case $\alpha$. In order to analyze the deformation by an operator of dimension $\Delta = 1$, we perform a Legendre transform by introducing the term \cite{Klebanov:1999tb,Papadimitriou:2007sj},
\begin{equation} \label{e:Sminus}
S_{-} = - \int_{\rho = \rho_{\text{cut}}} \D^3 x \sqrt{\gamma^\rho} \, \phi \,\pi_r\,,
\end{equation}
where $\pi_r$ denotes the renormalized canonical momentum:
\begin{equation}
\pi_r = \frac{1}{\sqrt{\gamma^\rho}} \frac{\delta (S_E + S_{\text{ct}})}{\delta \phi} = \partial_\rho \phi + \phi \,.
\end{equation}
This leads to the one-point function
\begin{equation}
\< \O_1 \>_s = \frac{1}{\sqrt{\gamma_{dS}}} \lim_{\rho \rightarrow \infty} \frac{\delta (S_E + S_{\text{ct}} + S_{-})}{\delta \beta} = \alpha\,.
\end{equation}

For future convenience, it is important to know the value of the renormalized on-shell action, $S_{\text{ren}}$, given by the sum of (\ref{e:SEU}, \ref{e:Sct}, \ref{e:Sminus}). Using the equations of motion and the asymptotic expansions, one can simplify this to:
\begin{equation} \label{SEren}
S_{\text{ren}} = \lim_{\rho_{\rm cut} \rightarrow \infty} \left[ - \int \D^4 x \sqrt{g_E} V(\phi) + \int_{\rho = \rho_{\text{cut}}} \D^3 x \sqrt{\gamma^\rho} \left( 2 + \frac{1}{2} R[\gamma^\rho] - K - \frac{1}{2} \phi^2 - \frac{1}{2} \frac{\D}{\D \rho} \phi^2 \right) \right]\,.
\end{equation}
The value of the on-shell action can easily be  calculated for empty $AdS$, for which $\A(\rho) = \sinh \rho$ and $\phi = 0$. In this case one finds $S_{\text{ren}} = \pi^2$.

\subsection{Free theory on de Sitter space} \label{sec:free}
We now return to Lorentzian signature and consider a conformally coupled massive scalar field $\varphi$ with the action
\begin{align} 
S &= - \int \D^3 x \sqrt{-\gamma_{dS}} \left[ \frac{1}{2} \partial_\mu \varphi \partial^\mu \varphi + \frac{1}{2} \xi R \varphi^2 + \beta \varphi^2 \right] \nonumber \\
&= - \int \D^3 x \sqrt{-\gamma_{dS}} \left[ \frac{1}{2} \partial_\mu \varphi \partial^\mu \varphi + \left( \frac{3}{8} +\beta \right) \varphi^2 \right],  \label{dSaction}
\end{align}
where in the second line we use the fact that the Ricci scalar $R = 6$ and the conformal coupling is given by $\xi = 1/8$. The bulk instanton geometries with fixed $\beta$ correspond to deforming the conformal case by adding a mass \cite{Maldacena:2010un}, $\beta = m^2/2$. Since canonically normalized scalars in 3 spacetime dimensions have dimension $1/2$, the deformation operator is: $ \O_1 =  \varphi^2$.

A massive field on de Sitter space is a well-defined, local QFT for any $\beta > -3/8$. Its two-point function $G = G(Z)$ is a function only of the de Sitter invariant $Z$, which is defined as an invariant element of the embedding space:
\begin{equation}\label{Zdef}
Z = \eta_{\mu\nu} \Delta X^\mu \Delta X^\nu\,.
\end{equation}
The embedding coordinates $X^\mu$, $\mu = 0,1,\ldots,d$ specify de Sitter space of unit radius by the equation $\eta_{\mu\nu} X^\mu X^\nu = 1$.  The geodesic distance, $D$, is related to $Z$ by
\begin{equation} \label{ZofD}
Z = \left\{ \begin{array}{ll} \cos D & \text{ for } D \leq \pi \,, \\
- \cosh(D - \pi) & \text{ for } D > \pi \,. \end{array} 
\right.
\end{equation}

\begin{figure}[b]
\begin{center}
\includegraphics[angle=0, width=0.7\textwidth]{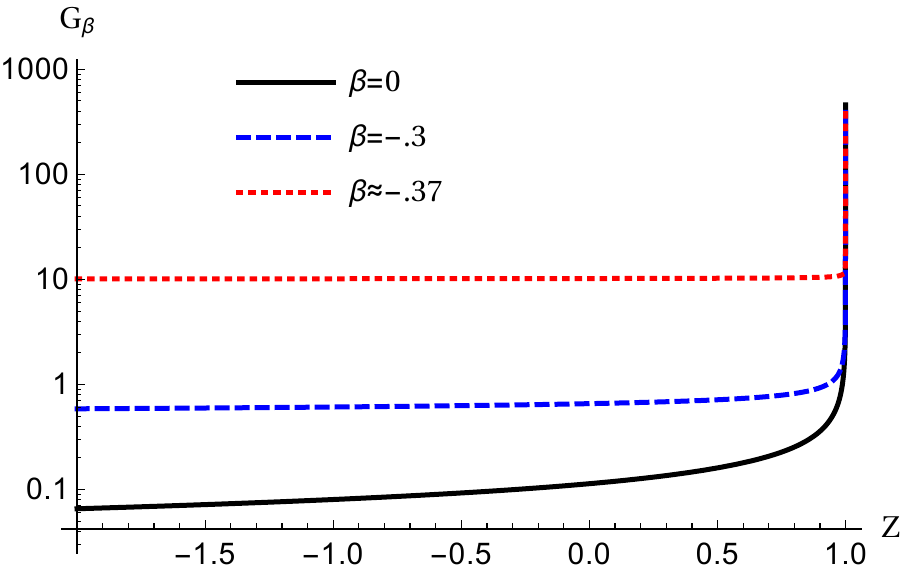}
\end{center}
\caption{The two-point function for a free, massive, conformally coupled scalar field in de Sitter space, as a function of the de Sitter invariant $Z$, for various values of the mass. As the mass-squared approaches the critical value $\beta_c^{fr} = m^2/2= -3/8$, where the scalar becomes massless, the two-point function develops a strong IR tail. In particular, it tends to a constant which diverges in the massless limit, with a subleading logarithmic decay.}
\label{scalar2pt}
\end{figure}

The two-point function is a solution to the equation of motion,
\begin{align} \label{dSKG}
0 &= \left[ - \Box + \left( \frac{3}{4} + 2 \beta \right) \right] G \nn\\
&= (Z^2 - 1) G''(Z) + 3 Z G'(Z) + \left( \frac{3}{4} + 2 \beta \right) G(Z)\,,
\end{align}
with the correct (properly normalized) short-distance divergence. In three dimensions, the solution can be written in terms of elementary functions,
\begin{equation} \label{Gfull}
G_\beta(Z) = \frac{\sin \left( \sqrt{1 - 8 \beta} \arcsin \sqrt{\frac{1 + Z}{2}} \right)}{2 \pi \sin \left( \frac{\pi}{2} \sqrt{1 - 8 \beta} \right) \sqrt{1 - Z^2}}\,.
\end{equation}
Its behavior for a range of masses is illustrated in \figref{scalar2pt}.

In the small distance limit, $Z \rightarrow 1^{-}$, the two-point function exhibits the correct singular behavior
\begin{equation} \label{shortdistdiv}
G_\beta(Z\to 1^-) = \frac{1}{4 \pi} \sqrt{\frac{2}{1 - Z}} + O((1 - Z)^0)\,.
\end{equation}
In the case of a massless, conformally coupled scalar, $\beta = 0$, the two-point function truncates to the single term written in \eqref{shortdistdiv}.

\paragraph{The massless limit}

The limit $\beta \to -3/8$ corresponds to the minimally coupled massless field as indicated by \eqref{dSaction}. This is a subtle case due to the fact that on the sphere, the defining equation
\begin{equation} \label{fundsoln}
- \Box_x G(x, x') = \sqrt{g(x')} \delta(x - x')
\end{equation}
has no solutions as its left hand side integrates to zero, while the right hand side does not. This problem manifests itself in various ways, such as the two-point function either breaking de Sitter invariance, not satisfying the equations of motion, or having unphysical divergences at finite separations, \cite{Allen:1985ux}. In particular, the naive limit $\lim_{\beta\to-3/8} G_\beta$ is ill-defined: one finds,
\begin{equation} \label{Gmasslesslim}
G_{\beta\to-3/8}(Z) = \frac{1}{2 \pi^2 \left( \frac{3}{8} + \beta \right)} + \frac{1}{4 \pi^2} \left[ -1 + \frac{4 Z \arcsin \sqrt{\frac{1+Z}{2}}}{\sqrt{1 - Z^2}} \right] + O\left( \frac{3}{8} + \beta \right).
\end{equation}

Various proposals to resolve this have been discussed in the literature. One suggestion is that the vacuum is only invariant under a subgroup of the full de Sitter symmetry group \cite{Allen:1985ux}.  A related possibility is to remove the zero-mode, which violates unitarity and is therefore considered unphysical \cite{Kirsten:1993ug, Tolley:2001gg}.  Additionally, one might restrict to considering only derivatives of the field in which divergences are absent; these can be used to describe all observable quantities. All of these amount to removing the divergent term in \eqref{Gmasslesslim}. Finally, it has been pointed out that the absence of a de Sitter invariant vacuum is a property of the free theory only, and is not an obstacle in interacting theories \cite{Einhorn:2002nu}.

\paragraph{Long range behavior}

We are particularly interested in the long range behavior of the two-point function. To examine this limit we first consider $-3/8<\beta <1/8$ and expand \eqref{Gfull} around $Z \rightarrow -\infty$, 
\be \label{lgdistfreescalar}
\begin{split}
G_\beta(Z\to -\infty) & = \({- (-2 Z)^{-\frac{1}{2} \left( 2 + \sqrt{1 - 8\beta} \right)} +  (-2 Z)^{-\frac{1}{2} \left( 2 - \sqrt{1 - 8\beta} \right)} \over 2 \pi \sin \left( \frac{\pi}{2} \sqrt{1 - 8\beta} \right)}\) \left[ 1 + O(Z^{-2}) \right].
\end{split}
\ee
Eq.~\eqref{lgdistfreescalar} shows that as $\beta$ approaches the critical value, $\bt_c^{fr} = -3/8$, where the scalar becomes massless, the two-point function develops a strong IR tail. Taking the naive limit $\beta \rightarrow -3/8$ using \eqref{Gmasslesslim} we find
\begin{equation} \label{e:Gm3/4}
G_{\beta\to-3/8}(Z\to -\infty) = \frac{1}{2 \pi^2 \left( \frac{3}{8} +\beta\right)} - \frac{\log(- Z) + 1 + 2 \log 2}{2 \pi^2} + O\left( \frac{3}{8} + \beta, Z^{-1} \right).
\end{equation}
The naive massless correlator therefore tends to a constant which diverges in the massless limit, plus a slowly decaying logarithmic tail.
We note that the constants in the second term of the above expression depend on the order of limits, and are furthermore irrelevant since one can always add an arbitrary constant to the fundamental solution of \eqref{fundsoln}.  On the other hand, the divergent first term and the logarithmic term in \eqref{e:Gm3/4} are physical signatures of the two-point function approaching the massless limit.

\paragraph{One-point function}

Since the bulk theory exhibits a non-trivial value of the one-point function, $\alpha =\< \O_1 \> \neq 0$, we would like to examine the one-point function of $\<\varphi^2\>$ in the free theory as a function of $\bt$. One can compute this by taking the zero-separation limit of the two-point function, $\<\varphi(x) \varphi(x') \>$. Since the two-point function is defined on the sphere, where IR divergences are absent, a natural procedure is to subtract the universal UV divergence \eqref{shortdistdiv}. This leads to the following definition of $\varphi^2$
\begin{equation} \label{e:1pt_from_2pt}
\varphi^2(x) = \lim_{x' \rightarrow x} \left[ \varphi(x) \varphi(x') - \< \varphi(x) \varphi(x') \>_0 \times 1 \right]\,,
\end{equation}
where $\< \varphi \varphi \>_0$ denotes the conformally coupled two-point function with $\beta=0$. This subtraction amounts to taking the constant part of the expansion \eqref{shortdistdiv}, given by:
\begin{equation} \label{vevfree}
\alpha=\< \varphi^2 \> = - \frac{\sqrt{1 - 8\beta}}{4 \pi} \cot \left( \frac{\pi}{2} \sqrt{1 -8\beta} \right).
\end{equation}

\begin{figure}[ht]
\begin{center}
\includegraphics[angle=0, width=0.7\textwidth]{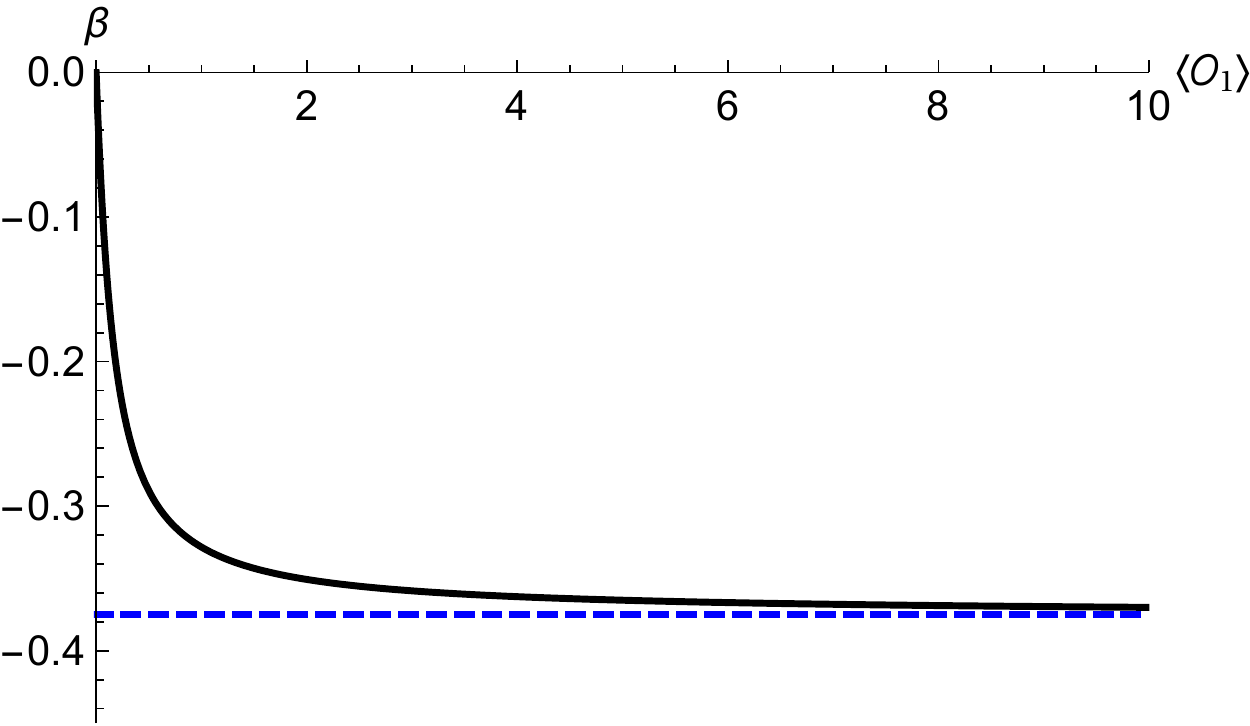}
\end{center}
\caption{The one-point function, $\<\varphi^2\>=\alpha$, as a function of the source, $\beta$. The one-point function diverges at a critical value, $\beta_c^{fr}=-3/8$, shown by the dashed line.}
\label{vevofsource}
\end{figure}

Notice that despite being a free theory, the one-point function is non-vanishing.  This is because the subtraction scheme \eqref{e:1pt_from_2pt} is $\beta$ independent.  In this way, the universal UV divergence is subtracted without modifying the IR.  The resulting non-zero one-point function is equal to that obtained from the generating functional for a free theory on the sphere \cite{Anninos:2012ft, Gabriele:2015gca}.  As $\beta$ approaches the critical deformation, $\beta\to -3/8$, the one-point function diverges. This behavior as a function of $\bt$ is illustrated in \figref{vevofsource} and closely resembles the behavior found at strong coupling using holography (\emph{cf.}~\figref{betaalpha} (left)).

\paragraph{Two-point function}

In Section \ref{georesults} we  study the strongly coupled ABJM theory by calculating the two-point functions of operators with $\Delta \gg 1$ via the geodesic approximation. As a point of comparison, we can model a two-point function of an operator $\O_{\Delta}$ of large dimension, $\Delta$, in the free massive theory by considering $\O_{N/2} = \lwick \varphi^{N} \rwick$. As is the case for the flat space theory, Wick's theorem is applicable and the two-point function reads
\begin{equation} \label{lgdim2pt}
\< \O_{N/2}(x) \O_{N/2}(x') \> = N! G^N(Z)\,,
\end{equation}
where  $G$ is the propagator given by \eqref{Gfull}.

\section{Phase transition at strong coupling} \label{DesignerGrav}

The free boundary theory is unstable for large negative deformations $\beta < \bt_c^{fr}=-3/8$, for which the boundary scalar $\varphi$ has a negative effective mass. The divergence of the one-point function, $\al$, at the critical deformation $\bt \rightarrow \bt^{fr}_c$ is a clear sign of the onset of this instability (\emph{cf.}~\figref{vevofsource}). In this section we identify and explore this phase transition at strong coupling in the boundary theory using holography.

We first construct an effective potential for the expectation value $\<\O_1\>$ in the dual field theory from the asymptotic behavior of the instantons encapsulated in the curve $\bt_i(\al)$, together with our choice of bulk boundary conditions. The construction of the effective potential for a dual field theory on de Sitter space is a generalization of the results of \cite{Hertog:2004ns}, where effective potentials for dual theories defined on the global $AdS$ boundary were calculated using the asymptotic behavior of static scalar soliton solutions. Static solitons can be vacua of the theory for mixed boundary conditions, $\bt(\al)$, in global $AdS$ \cite{Hertog:2004ns}; similarly the instantons discussed in Section \ref{setup} can be vacua for the theory with non-trivial boundary conditions on the de Sitter boundary.

Specifically, we consider the following function of $\alpha = \<\O_1\>$,
\be \label{e:Veff}
V_{\rm eff}(\alpha) = -\int_0^\alpha \beta_{i}(\alpha) d\alpha + \int_0^\alpha \beta(\alpha) \D \alpha,
\ee
where $\beta_{i}(\alpha)$ is shown in \figref{betaalpha} (left) for strong coupling, and \figref{vevofsource} for the free theory.  In the second term of \eqref{e:Veff} $\beta(\alpha)$ refers to the boundary conditions on the de Sitter boundary and with our choice of von Neumann boundary conditions $\beta(\alpha) = \beta$.

We now argue that the function \eqref{e:Veff} is equal to the standard effective potential. In a QFT the quantum effective action solves the quantum equation of motion, which in our notation reads
\begin{equation} \label{gamma0}
\frac{\delta \Gamma_0}{\delta \alpha(x)} =  \sqrt{-\gamma_{\rm dS}} \beta(x)\,,
\end{equation}
where $\Gamma_0$ is the quantum effective action of the undeformed theory, and $\alpha$ is the vacuum one-point function in the presence of the source $\bt$. The quantum effective action of the deformed theory, $\Gamma_\beta$, is then given by
\begin{equation}
\Gamma_\beta(\alpha) = \Gamma_0(\alpha) - \int d^3x \sqrt{-\gamma_{\rm dS}} \int_0^{\alpha(x)} \beta(\alpha) \D \alpha\,.
\end{equation}
The relation between the quantum effective action and the effective potential in the case of constant $\alpha$ is $\Gamma_\beta = -Vol_{\rm dS} V_{\rm eff}$.  Using \eqref{e:Veff} implies $\Gamma_0=-Vol_{\rm dS} \int \beta_{i} d\alpha$, and we see that \eqref{gamma0} is satisfied for the stationary solutions, $\beta=\beta_{i}$. Therefore \eqref{e:Veff} plays the role of an effective potential for $\al$. In particular, its extrema are in one-to-one correspondence with the regular instantons that obey the boundary conditions $\beta(\alpha)$. 

\begin{figure}[ht] 
\begin{center}$
\begin{array}{|c | c| }
\hline
\includegraphics[angle=0, width=0.48\textwidth]{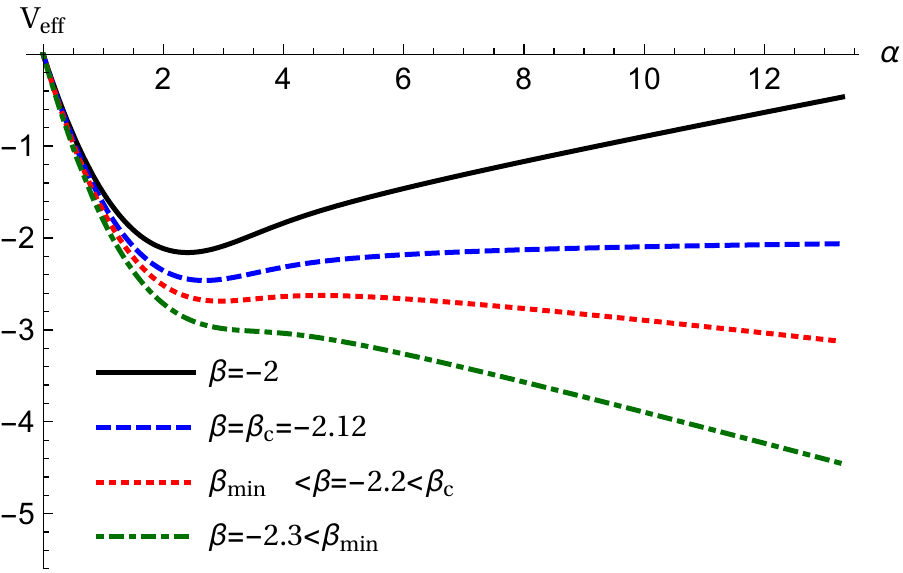} & \includegraphics[angle=0, width=0.48\textwidth]{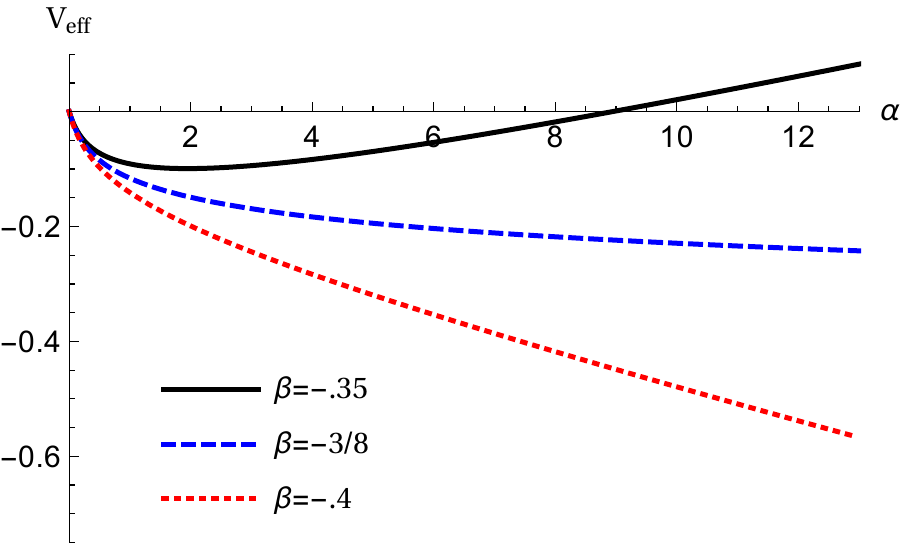} \\
\hline
\end{array}$
\end{center}
\caption{Left panel: Effective potential for the source $\alpha = \<\varphi^2\>$ calculated for the ABJM boundary theory dual to a crunching cosmology.  Right panel: The analogous effective potential for $\<\varphi^2\>$ in the theory of a conformally coupled massive scalar on de Sitter, where $m^2$ is the deformation.}
\label{fig:Veff}
\end{figure}

The effective potential \eqref{e:Veff} is shown for strong coupling in \figref{fig:Veff} (left). For small deformations $V_{\rm eff}$ has a stable minimum which indicates that the strongly coupled boundary theory remains stable just like the free theory. By contrast, for large negative deformations $\beta<\beta_{\rm min}$ we see $V_{\rm eff}$ has no extrema at all, and the boundary theory is clearly unstable. Finally there is an interval $\beta_{\rm min}<\beta\le \beta_c$ in which our analysis of the effective potential indicates the theory is metastable. In this regime $V_{\rm eff}$ has two extrema which correspond to two different instanton solutions\footnote{This can also be seen in \figref{betaalpha} (left).}. The `second' instanton, with the larger value of $\al$, corresponds to a maximum of the effective potential and is presumably unstable. The first instanton describes a perturbatively stable vacuum state.

A genuine effective potential in a stable QFT is convex. Clearly this is not the case for $\beta \le \beta_c$ in \figref{fig:Veff} (left). 
The reason is that we consider a state where the one-point function, $\< \O_1 \>$, is defined to be position-independent in a regime where  the vacuum is at best metastable. In this regime one expects there to be a spatially varying state in the QFT with a lower value of the effective potential. For $\beta_{\min} < \beta < \beta_c$ such states correspond to instanton solutions in the dual field theory that tunnel away from the perturbatively stable minimum. Hence our interpretation of \eqref{e:Veff} as a genuine effective potential for $\< \O_1 \>$ allows us to conclude that for $\beta_{\min} < \beta < \beta_c$ the field theory becomes non-perturbatively unstable.

\figref{fig:Veff} (right) shows the effective potential for the operator $\<\varphi^2\>$ in the theory of a free, massive, conformally coupled scalar in de Sitter space discussed in Section \ref{sec:free}. This effective potential is constructed by inverting \eqref{vevfree}, substituting it in \eqref{e:Veff} for $\beta_i(\al)$, and integrating.  The mass deformation away from the conformal point is given by the deformation parameter $\beta=m^2/2$. A comparison between both panels of \figref{fig:Veff} shows that the transition to the unstable regime for large (negative) deformations occurs in a qualitatively similar manner. On the other hand, the critical value of $\bt_c$ at which the instability sets in is more negative at strong coupling. Also, strong coupling allows for a narrow metastable regime which is absent in the free theory.

The most important feature of this analysis, which we will use below, is that both regimes exhibit a critical deformation, $\bt_c$, at which the theory becomes unstable. In the free theory this transition corresponds to the case of a massless, minimally coupled scalar in de Sitter. In the next Section we show that the near critical regime allows us to probe the high curvature region near the singularity with boundary observables.

\section{Holographic signatures of the singularity} \label{georesults}
In the large $N$ limit of the field theory, the leading contribution to the two-point correlator of an operator $\mathcal{O}$ of high conformal dimension $\Delta$ in the dual strongly coupled ABJM theory on the de Sitter boundary is given in terms of the length of spacelike bulk geodesics connecting the two points:
\cite{Balasubramanian:1999zv}
\be \label{holodict}
\langle \psi | \O_\Delta(x)\O_\Delta(x') | \psi \rangle = \sum_i w_i \e^{-\Delta \mathcal{L}^i_{\text{reg}} (x,x')}\,,
\ee
where $|\psi\rangle$ is the state of the boundary theory, $\mathcal{L}_{\text{reg}} (x,x')$ is the regulated length of the geodesics connecting boundary points $x$ and $x'$, and $w_i$ are the relative weights of the contributions of different saddle point geometries. We work in a state where the weights, $w_i$, are related to the Euclidean on-shell action evaluated on the bulk backgrounds.

The calculation of the geodesics and their lengths is presented in full detail in Appendices \ref{secgeos} and \ref{seclengths}, here we summarise the key points and discuss the resulting correlator. Geodesics in $AdS$ cosmologies of this kind have also been recently studied in \cite{Kumar:2015jxa}, and many of the results in the Appendices have overlap with this work.

\subsection{Bulk geodesics} \label{geolentext}
The de Sitter symmetry of the boundary implies the correlator can only depend on the de Sitter invariant distance between $x$ and $x'$, given in eqs. \eqref{Zdef} and \eqref{ZofD}. We  therefore restrict attention to geodesics connecting boundary points at equal times, $\tau_B$. For sub-horizon boundary separations, equal time geodesics are symmetric around a turning point that is located outside the lightcone emanating from the origin at $\tau=0$ (\emph{cf.}~\figref{crunchgeo}). Such geodesics do not probe the recollapsing FLRW patch of the bulk. The antipodal geodesic with $\tau_B=0$ has a boundary separation equal to the de Sitter horizon. The time-symmetry of the initial data implies this geodesic passes through the tip of the lightcone bounding the interior region. Larger, super-horizon boundary separations can be covered by considering antipodal bulk geodesics with increasing $\tau_B$. These are symmetric around a turning point, $t_{\rm turn}$, at the center ($\chi=0$) of a spatial slice inside the lightcone.

This one-parameter set of highly symmetric geodesics covering all boundary separations depends on a single integration constant $E$, which is defined in eq. \eqref{e:efin2} and related to the scale factor at the turning point in the bulk as follows,
\begin{equation} \label{e:bendtext}
a(t_{\rm turn}) = \sqrt{\frac{1}{E}}\,.
\end{equation}
Geodesics connecting sub-horizon boundary separations turn around outside the interior FLRW patch at a value $\rho_{\rm turn}=-i t_{\rm turn}$ and hence correspond to negative values of $E$. The $E\to 0^-$ limit corresponds to the short distance limit in which the geodesics only probe the asymptotic $AdS$ geometry, while the $E\to-\infty$ limit corresponds to the antipodal geodesic that passes through the tip of the lightcone emanating from the origin. Geodesics connecting super-horizon boundary separations have positive $E>E_m>1$, where the $E \to  E_m$ limit corresponds to the limit in which the turning point $t_{\rm turn} \to t_{\rm max}$. Recall that $t_{\rm max}$ is the time at which the FLRW scale factor is maximized. This is the limit in which the geodesics come closest to the singularity in the interior cosmological patch. Larger positive values of $E$ correspond to geodesics turning around at smaller values of the scale factor $a$ and stay further away from the singularity. There are no geodesics anchored on the boundary that enter the region inside the cosmological patch beyond $a_{\rm max}$ \cite{Engelhardt:2013tra}. 

The relation between $E$ and the boundary separation $Z$ is given by (\emph{cf.}~\secref{app:dSdist}),
\be \label{Zoftturntext}
Z=\cosh\( 2 \int_{\I \infty}^{t_{\rm turn}(E)} {dx \over a(x)\sqrt{1-Ea^2(x)}} \)\,.
\ee
Since we want to probe the region of the bulk near the singularity, we are particularly interested in the $t_{\rm turn} \to t_{\rm max}$ limit. Expanding \eqref{Zoftturntext} around $t_{\rm max}-t_{\rm turn}$ we get, in terms of the de Sitter invariant distance $D$, 
\be
D = \frac{-1}{\sqrt{-a_{\rm max} a''_{\rm max}}} \log (t_{\rm max} - t_{\turn})^2 + \mathcal{O} \left((t_{\rm max} - t_{\turn})^0 \right)\,,
\ee
where $a_{\rm max}''\equiv a''(t)|_{t=t_{\rm max}}$. This shows that as we increase $D$ the geodesics `pile up' near the surface $a(t)=a_{\rm max}$. We therefore expect the clearest signals of the singularity in the large distance limit of the boundary correlator, in line with the usual UV/IR correspondence of $AdS$/CFT. The length of the geodesics is given by (\emph{cf.}~\eqref{lenregfin}):
\be \label{lenregfintext}
\L = 2\int_{\rho_{\rm cut}}^{t_{\rm turn}} {\sqrt{E}a(t) dt \over \sqrt{1-Ea^2(t)}}\,,
\ee
where we have introduced a cut-off, $\rho_{\rm cut}=it_{\rm cut}$, which we specify in more detail below. Expanding \eqref{lenregfintext} in the near boundary regime leads to  \eqref{lregZsm}:
\be \label{Lshort}
\mathcal{L} (Z\to 1^-)  = -2\log\epsilon+ \log \(2(1-Z) \) -{\alpha^2\over 12}(1-Z)+\mathcal{O}(1-Z)^2\,, 
\ee
where $\alpha$ is given by the asymptotic expansion of the scalar field \eqref{aysphi}, and the cut-off $1/\epsilon \equiv a_{\rm out}(\rho_{\rm cut})$. 
As usual we regulate the length by introducing a cut-off at a fixed large value of the conformal factor, and subtracting the universal contribution. This yields:
\be \label{renscheme}
\begin{split}
\L_{\rm reg} &= \lim_{\epsilon \to 0} \( \L +2\log \epsilon \) \\
&= \lim_{\rho_{\rm cut}\to \infty} \(\L -2\rho_{\rm cut} \) -\log(a_1^2) \, ,\\ 
\end{split}
\ee	
where we use the expansion of the scale factor \eqref{aysa}  in the second line.
Using this scheme, the correlators \eqref{holodict} exhibit the correct, universal short distance behavior\footnote{The normalization constant can be chosen independently.} :
\be \label{shortDCFT}
\< \O_\Delta(x) \O_\Delta(x') \> \propto \( {2\over1-Z} \)^{\Delta} + \text{ subleading terms}\,,
\ee
in agreement with the weak field behavior \eqref{shortdistdiv}.
\begin{figure}[ht] 
\begin{center}$
\begin{array}{|c | c| }
\hline
\includegraphics[angle=0, width=0.48\textwidth]{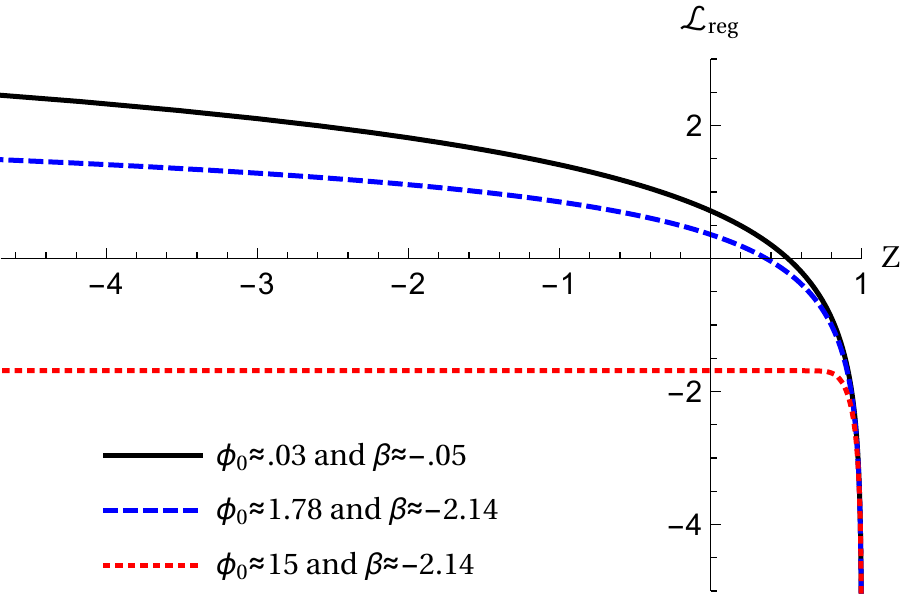} & \includegraphics[angle=0, width=0.48\textwidth]{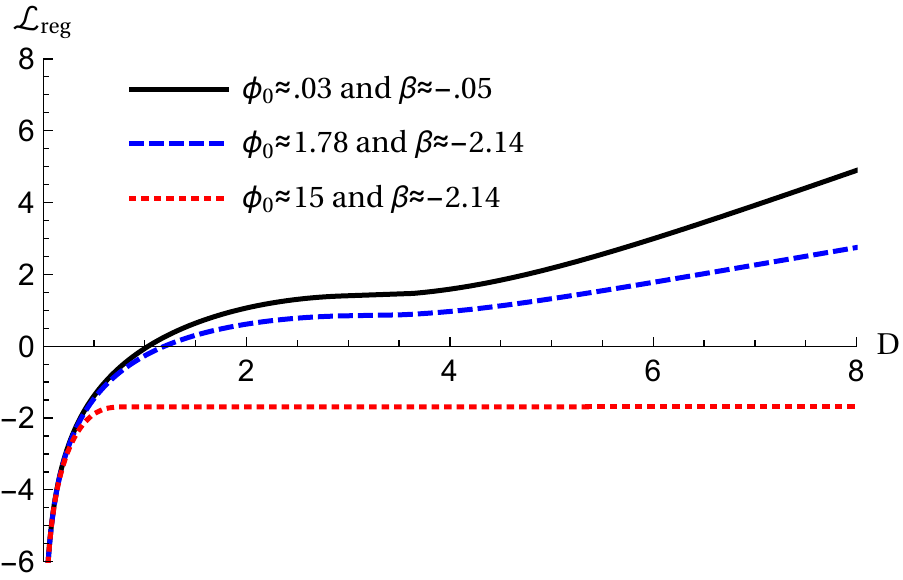} \\
\hline
\end{array}$
\end{center}
\caption{The regulated length of geodesics as a function of the de Sitter invariant $Z$ (left) and as a function of the de Sitter invariant distance $D$ (right), for several backgrounds.}
\label{fig:lengeo}
\end{figure}

The regulated length as a function of the boundary separation is shown in \figref{fig:lengeo} in three different backgrounds. One sees that $\L_{\rm reg}$ increases more slowly with the boundary separation in backgrounds with larger $\phi_0$, the value of the scalar field on the lightcone emanating from the origin. This is because the singularity tends more and more to a light-like singularity as $\phi_0$ increases. This has the effect that for large values of $\phi_0$ the bulk geodesics in the interior region also become more light-like, so their length grows more slowly. Since our regularization scheme \eqref{renscheme} is universal, this leads to a smaller regulated length $\L_{\rm reg}$ in the large distance regime for increasing $\phi_0$.

\subsection{Holographic two-point functions}
We now use our results for $\L_{\rm reg}(D)$ to evaluate the boundary two-point function \eqref{holodict}. We have seen that for a given boundary separation there is a unique geodesic in each bulk background that connects the two boundary points. But there may be more than one background that contributes to the boundary correlator! Each instanton background provides a saddle point contribution in the holographic calculation of the boundary correlator \eqref{holodict}. The bulk boundary conditions, $\beta$, determine the dual theory and therefore all instantons with a given $\beta$ contribute to the correlator. This is in contrast to previous geodesic probes of singularities (\emph{e.g.}, \cite{Fidkowski:2003nf, Engelhardt:2014mea}), where multiple geodesics in a single background contribute to the correlator.

\figref{betaalpha} shows that for $\beta >\beta_c$ there is only one background, so the relative weighting factor in \eqref{holodict} is irrelevant in this regime. However, boundary conditions $\beta_{\min}\le \beta \le \beta_c $ admit two regular backgrounds and the question arises what are their relative weights in the gravitational path integral that leads to \eqref{holodict}. This is where our Euclidean initial conditions enter; they correspond to starting the bulk in the vacuum state for $\bt \neq 0$ boundary conditions, and provide a natural relative weighing given by $w_i \propto \exp(-S_E)$, where $S_E$ are the renormalized Euclidean actions of the instantons \eqref{SEren}.
To summarize, we adopt the following relative weighting, 
\be \label{weights}
w_i = \left\{ \begin{array}{ll} 1 & \text{ for } \beta > \beta_c\,, \\
{\e^{-S_{E \, i}} \over \sum_j \e^{-S_{E \, j}}} & \text{ for } \beta_{\min}\le \beta \le \beta_c \,, \end{array} 
\right.
\end{equation}
where the sum runs over backgrounds with $\beta_{i}(\alpha)=\beta$. 

\begin{figure}[ht]
\begin{center}
\includegraphics[angle=0, width=0.7\textwidth]{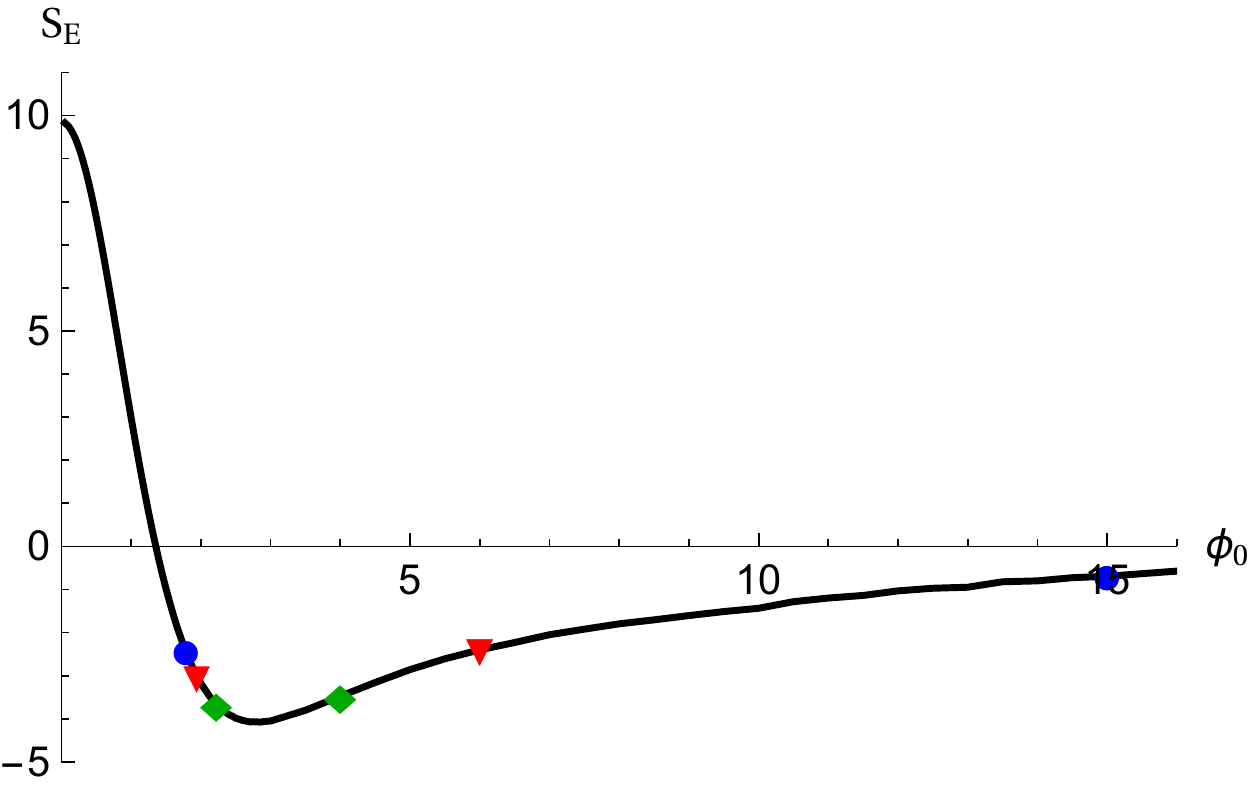}
\end{center}
\caption{The Euclidean action of the instantons as a function of $\phi_0$.  Matching symbols mark pairs with the same value of $\beta$.}
\label{SE}
\end{figure}
%
We plot the renormalized Euclidean actions \eqref{SEren} of the instantons in \figref{SE}, as a function of $\phi_0$. This shows that for boundary conditions that admit two instantons, the saddle point with the largest value of $\phi_0$ is always sub-dominant relative to the small $\phi_0$ saddle point. 

\begin{figure}[ht]
\begin{center}
\includegraphics[angle=0, width=0.7\textwidth]{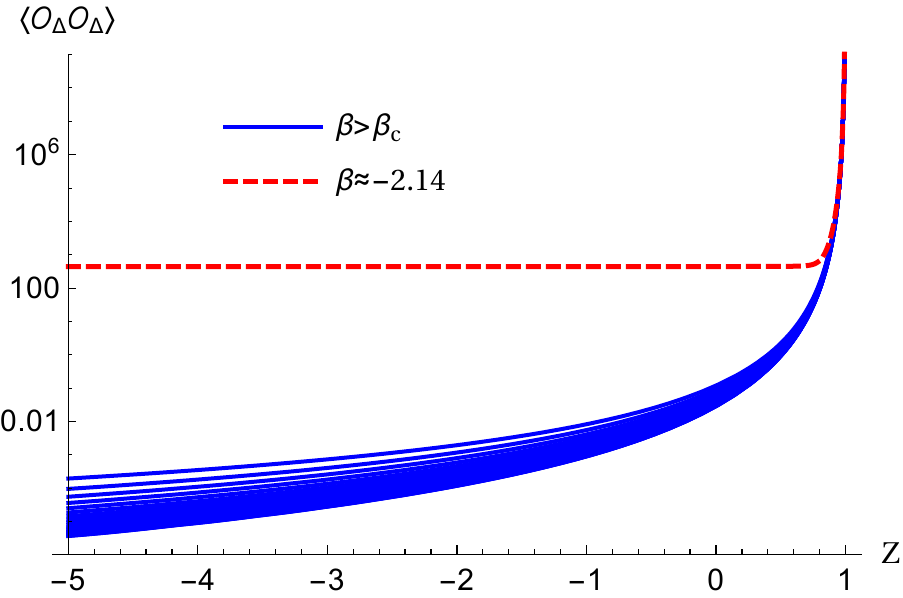}
\end{center}
\caption{The boundary two-point function for a large dimension operator computed holographically for a range of different deformations $\bt$. A phase transition occurs as $\beta \to \bt_c \sim -2.12$ where the two-point function develops a strong IR tail due to the contribution from a second saddle point geometry in which the spacelike geodesics probe the high curvature region near the singularity. Therefore we interpret this a signature of the singularity. We find a qualitatively similar enhancement of IR correlations near the critical point in the free theory, although there the transition is smooth. To be specific we take $\Delta=5$ in this figure. Also, the overall normalization is chosen so that the short-distance behavior given in \eqref{shortDCFT} is universal with the same constant of proportionality.}
\label{2ptbulk}
\end{figure}

Substituting the weights \eqref{weights} along with our results for the regulated lengths, summarised in \figref{fig:lengeo}, into \eqref{holodict} yields the two point correlation function for boundary operators with $\Delta \gg 1$. The results are shown in \figref{2ptbulk} for an operator with $\Delta=5$ for a range of different deformations $\bt$.

We are especially interested in the IR behavior of the correlator. This can be obtained analytically by expanding \eqref{Zoftturntext} and \eqref{lenregfintext} for large boundary separation and applying our regularization scheme \eqref{renscheme}. This yields, in the limit $Z \to -\infty$ (see also Appendix \ref{seclengths}):
\be
\mathcal{L}_{\rm reg} = a_{\rm max}(\beta) \log \(-2Z \)+ \mathcal{O}(Z^0)\,. \label{Llong}
\ee
Substituting \eqref{Llong} in \eqref{holodict}, with $w_i$ given by \eqref{weights}, then gives:
\be \label{geolg}
\< \O_\Delta(x) \O_\Delta(x') \> = \sum_i w_i\, \left(C(\beta) (-2Z)^{- a_{{\rm max}}^{i} (\beta) } \right)^{\Delta} +\text{  subleading terms for } Z\to -\infty\,,  
\ee
where $C(\beta)$ represents the constant terms in \eqref{Llong}. 
We derive an approximate analytic expression for $C(\beta)$ in \eqref{Llinear}\footnote{Note that this approximation becomes better with increasing $\phi_0$}. Substituting this in \eqref{geolg} yields for the large distance correlator,
\be \label{geolg2}
\< \O_\Delta(x) \O_\Delta(x') \>
\approx \sum_i w_i\, \left((a_1^i)^2  \e^{\pi a_{{\rm max}}^i} (-2Z)^{-a_{{\rm max}}^i}\right)^{\Delta} \,.
\ee

Thus, in the near critical regime where two saddle points contribute to the boundary correlator, the large $\al$ instanton, which has the smallest amplitude, nevertheless dominates the correlator in the IR. By contrast, it follows from \eqref{shortDCFT} and the relative weighting \eqref{weights} that the large $\al$ instanton provides a sub-dominant contribution to the short distance correlator. Its contribution at large distances leads to a strong enhancement of the correlations, a feature which we interpret as a clear signature of the singularity. Evidence for this interpretation comes form the fact that the geodesics come very close to the singularity in the large $\al$ backgrounds (\emph{cf.}~\eqref{propDtosing}). In fact, in the limit $\beta \to \beta_c$ the large $\al$ instantons become singular, with $\phi_0\to \infty$ and hence $a_{\rm max} \to 0$, corresponding in the Lorentzian to a light-like singularity emanating from the origin.
In the next section we explore the large distance behavior of the two-point function in the $\bt \sim \bt_c$ regime in more detail and in particular compare this with the two-point function in the free theory computed in Section \ref{CFTsec}.

\section{Comparison of weak and strong coupling}

We now compare the two-point function in the mass deformed ABJM theory computed holographically at strong coupling, with the two-point function of a high dimension operator in the theory of a free, massive scalar in de Sitter space. Both theories have a critical deformation, $\bt_c$, at which the theory undergoes a phase transition and becomes unstable. Moreover in both cases the expectation value of the deformation operator, $\<\O_1\>$ diverges at the critical deformation. It is therefore interesting to compare the behavior of the two-point functions in the neighbourhood of $\bt_c$. Since this is a comparison between very different regimes of the theory, we only expect a qualitatively similar behavior at best. 

We concentrate on the physically interesting IR regime. The large distance correlator at strong coupling is shown in \figref{2ptbulk} and given in \eqref{geolg2}. The leading behavior at large distance in the free theory is given by the leading (second) term of \eqref{lgdistfreescalar}, raised to the power $2\Delta$ according to \eqref{lgdim2pt}:
\be \label{freelg}
\< \O_\Delta(x) \O_\Delta(x') \>_{\rm free} = (2\Delta)! {(-2Z)^{-\Delta(2 - \sqrt{1-8\beta})} \over \(2\pi \sin \({\pi \over 2}\sqrt{1-8\beta}\)\)^{2\Delta}}+\text{  subleading terms for } Z\to -\infty\,.
\ee  
We now compare this to the strongly coupled results first for small $\bt$ - the `near $AdS$' limit - and then in the $\bt \sim \bt_c$ regime near the onset of the instability.

\paragraph{The $AdS$ limit:}
In the limit $\phi_0 \ll 1$ the bulk system can be solved perturbatively in $\bt$,
\be
 a_1 = {1\over 2} + {2\beta^2 \over 9}+\O(\beta^3)\,, \qquad a_{\rm max} = 1 - {4\beta^2 \over 3}+\O(\beta^3)\,.
\ee
Plugging these expressions into \eqref{geolg2} and expanding in small $\beta$ gives
\be
\begin{split}
\< \O_\Delta(x) \O_\Delta(x') \>_{\rm s.c.} &= \({\e^\pi \over 4}\)^{\Delta}\left[(-2Z)^{-\Delta} + {\Delta\over 12}(-2Z)^{-\Delta}\(\log(-2Z)+\pi-2/3\) \beta^2\right]\\
& \quad + \O(\beta^3)+\text{subleading terms for }  Z\to -\infty\,,
\end{split}
\ee
where the subscript `s.c.' refers to the strong coupling regime of the ABJM theory. 
On the other hand a small $\beta$ expansion of the free correlator \eqref{freelg} yields:
\be 
\begin{split}
\< \O_\Delta(x) \O_\Delta(x') \>_{\rm free} & = {(2\Delta)! \over (2\pi)^{2\Delta}}\left[(-2Z)^{-\Delta} - 4\Delta(-2Z)^{-\Delta}\log(-2Z) \beta\right]+\O(\beta^2) \\ & \quad +\text{  subleading terms for }  Z\to -\infty\,.
\end{split}
\ee
The leading term in both cases is that of a conformally coupled field, as it should be in the near $AdS$ limit. The first order in $\beta$ corrections have the same $Z$ dependence, but they enter at different orders in $\beta$ for weak versus strong coupling. 

\paragraph{The $\beta \to \beta_c$ limit:}
The critical deformation in the free theory is given by $\beta^{fr}_c=-3/8$. At this value the free scalar becomes massless and the two-point function develops a strong IR tail. In particular, \eqref{e:Gm3/4} raised to the $2\Delta$ gives:
\be \label{betaclimit}
\begin{split}
\< \O_\Delta(x) \O_\Delta(x') \>_{\rm free} & \propto {1 \over (\beta^{fr}_c-\beta)^{2\Delta}}\left[ 1 + 2 \Delta (\beta_c^{fr}-\beta)  \log(-2Z) +\O(\beta_c^{fr}-\beta)^2 \right]\, \\
& \quad +\text{  subleading terms for }  Z\to -\infty\,.
\end{split}
\ee

In the strongly coupled theory, the critical deformation occurs at $\beta=\beta_c \approx -2.12$. We have seen that the two-point function changes discontinuously when $\beta \to \beta_c^+$, due to the sudden appearance of a second saddle point. However we can also approach the transition point from below. In this way the transition is smooth and we can meaningfully expand the correlator in the small parameter $(\beta_c - \beta)>0$ and see how the IR tail emerges.

In the absence of an analytic approach to calculate the dependence of the various constants in \eqref{geolg2} on $\beta$, we turn to numerical fits. We fit $a_1$ and $a_{\rm max}$ in the regime $\phi_0 \gg 1$ corresponding to $\vert \beta - \beta_c\vert \ll 1$, which is the asymptotic tail of the functions in \figref{betaalpha}.  We find:
\be \label{numfits}
a_1 \approx {A \over \sqrt{\beta_c - \beta}}\, \qquad \log(a_{\rm max}) \approx B_1 - {B_2 \over \beta_c-\beta}\,,
\ee
with $A = 0.36$, $B_1 =0.61$, and $B_2= 0.26$. 
These expressions can be used in \eqref{geolg} to find the IR expansion around $\beta-\beta_c$,
\be \label{betaclimit2}
\begin{split}
\< \O_\Delta(x) \O_\Delta(x') \>_{\rm s.c.} &\propto  {1\over (\beta_c-\beta)^{\Delta}} \left[ 1 - \Delta \e^{B_1} \e^{-{B_2  \over \beta_c-\beta}} \log(-2Z) +\O\( \e^{-{2B_2 \over \beta_c-\beta}}\)\right] \\
& \quad +\text{  subleading terms for }  Z\to -\infty\,,
\end{split}
\ee
where the sub-dominant saddle point, with the larger value of $a_{\rm max}$, contributes only in the subleading terms. This reveals again a remarkable qualitatively similar behavior to the free field correlator \eqref{betaclimit} near the critical deformation.  Note that the difference in the  $\Delta$ dependence of \eqref{betaclimit} and \eqref{betaclimit2} is not surprising, as under RG flow the dimensions of operators typically vary.

It is furthermore interesting to notice that the expansion in $\exp\(-1/ (\beta_c-\beta)\)$ in \eqref{betaclimit2} is reminiscent of an instanton expansion around the massless theory.  This is especially suggestive when taken with the observation of \secref{DesignerGrav} that the boundary theory at strong coupling is non-perturbatively unstable in the regime $\beta \lesssim \beta_c$.  It would be interesting to explore this in more detail.

\section{Conclusions}

We have applied gauge/gravity duality to study the dynamics near cosmological singularities produced in scalar field driven cosmologies with asymptotic $AdS_4$ boundary conditions. The dual description of these $AdS$ cosmologies consists of a mass deformation of ABJM theory defined on de Sitter space.  This dual theory is well-defined and stable for sufficiently small deformations. 

We have identified a critical deformation where the theory becomes unstable. It is exactly in this near critical regime that the bulk singularity can be probed by boundary observables. Specifically, close to the critical point, spacelike geodesics with endpoints anchored at large boundary separation explore the high curvature region near the singularity. We have seen this strongly enhances IR correlations in the two-point function of a high dimension operator at strong coupling. At the critical point itself, spacelike geodesics with endpoints far apart on the boundary nearly touch the singularity in the bulk, leading to an IR divergence of the boundary correlator. This lends further support to the general picture that holographic signatures of bulk singularities appear to be encoded in the long-wavelength features of boundary observables.

This IR behavior of the two-point function at strong coupling has a natural interpretation in the dual, where in the free theory, the deformation operator at the critical point casuses a boundary scalar to become massless. The strong IR tail of the two-point function of a massless field in de Sitter is very similar to what we find at strong coupling. It would be very interesting to study whether and how the regulation of the correlator of a massless field in de Sitter is connected to the boundary conditions at the bulk singularity.

For deformations somewhat larger than the critical deformation there is a second bulk background that contributes to the boundary two-point function in the geodesic approximation. This is  reminiscent of the situation for black holes in $AdS$ \cite{Maldacena:2001kr}. There and elsewhere it has been argued that in $AdS$/CFT one should sum over bulk geometries in the Euclidean theory and include the thermal $AdS$ saddle point \emph{e.g.}, to restore unitarity. Since  we view the Euclidean path integral as defining our initial wave function, we are naturally led to sum over both saddle points, which then implies we must include both contributions in our Lorentzian computation of the correlator. As in the case of black holes, the contribution of the sub-dominant saddle point becomes relevant only in the large distance regime of the correlator. This is a fascinating first step towards a holographic understanding of the quantum nature of bulk singularity in terms of a superposition of two classical geometries.

\section*{Acknowledgements}
We thank Nikolay Bobev, Gabriele Conti, Netta Engelhardt, Fri{\dh}rik Gautason, Alessandra Gnecchi, Gary Horowitz and Tom Lemmens for useful conversations.  We thank the KITP and the Physics Department at UCSB for their hospitality. The work of AB and TH is supported in part by the Belgian National Science Foundation (FWO) grant G.001.12 Odysseus and by the European Research Council grant no. ERC-2013-CoG 616732 HoloQosmos. The work of MS is supported by funding from the European Union's Horizon 2020 research and innovation
programme under the Marie Sk{\l}odowska-Curie grant agreement No 656491.


\appendix
\section{Quantifying the crunch} \label{app:crunch}
This appendix contains a few brief results which elucidate some features of our bulk solutions producing a big crunch. The crunch refers to the fact that, in the interior of the lightcone emanating from the origin, at a finite time $t_{c}$, the scale factor vanishes, $a(t_{c}) = 0$ (\figref{crunchgeo}). This does not automatically imply that the spacetime becomes singular, \emph{e.g.}, in empty $AdS$, $a(t) = \sin t$, the vanishing of the scale factor at $t_{c} = \pi$ is merely a coordinate singularity.  We will show that for any non-trivial scalar profile ($\phi(\rho) \neq \text{const.}$), there is a true singularity at finite $t_{c}$, \emph{i.e.}, only the empty $AdS$ example avoids the presence of a singularity.

The regularity conditions \eqref{e:regularity} imply that the scale factor grows for some range $0 < t < t_{\rm max}$, attaining its maximum $a_{\rm max}$ at some time $t_{\rm max}$. However there is an upper bound on the scale factor given by the $AdS$ solution,
\begin{equation} \label{e:abound}
a(t) < R\sin\( t/R \) < \sin t~, \qquad \text{where  } R=\sqrt{{3\over |V(\phi_0)|}}\,,
\end{equation}
where $\phi(\rho=0) \equiv \phi_0$, which in turn implies 
\be \label{amaxbound}
t_{\rm max} \leq R\pi/2 < \pi/2\,, \qquad  a_{\rm max} \leq R < 1 \,.
\ee
To find the bound \eqref{e:abound} we first subtract twice \eqref{e:fr2} from \eqref{e:fr1}:
\begin{equation} \label{e:ein1}
\frac{a'^2 - 1}{a^2} = {1\over 3}\left({1\over 2} \phi'^2 + V(\phi)\right),
\end{equation}
When $t < t_{\rm max}$ we have, by definition, $a'(t) > 0$ and \eqref{e:kg} implies:
\begin{equation}
\phi'( \phi'' + V'(\phi) ) = - 3 \frac{a'}{a} \phi'^2 < 0.
\end{equation}
Therefore:
\be
{d \over dt} \( \frac{a'^2 - 1}{a^2} \) = {1\over 3}\phi'\left( \phi'' + V'(\phi)\right)<0.
\ee
Thus, the left hand side of \eqref{e:ein1} decreases monotonically with $t$ for $t<t_{\rm max}$, and is therefore bounded by its value at $t = 0$,
\begin{equation} \label{ineqfin}
\frac{a'^2 - 1}{a^2} < {1\over 3} V(\phi_0) < -1,
\end{equation}
where we have used the fact that $\phi'(t=0)=0$ \eqref{e:regularity}. The expression \eqref{ineqfin} integrates to \eqref{e:abound} for $t < t_{\rm max}$.

Similarly, one can show that the value of the scalar field must grow monotonically  inside  the lightcone. Indeed, if $\phi$ has a maximum at some $t_0 > 0$, then $\phi'(t_0) = 0$ and $\phi''(t_0) \leq 0$. The Klein-Gordon equation \eqref{e:kg} then implies $V'(\phi(t_0)) \geq 0$, which for our potential \eqref{potential} can only be satisfied by saturation at $\phi(t_0) = 0$.

Finally, we argue that if the singularity at $t_{c}$ is a coordinate singularity, then the only regular background geometry is empty $AdS$. Note that if $\phi'(t_{c})$ remains finite, then equation \eqref{e:fr1} implies that $a'(t_{c}) = -1$. From \eqref{e:abound}, we see that this cannot be satisfied for non trivial scalar profiles (\emph{i.e.}, it is only satisfied by pure $AdS$). Hence the only non-trivial solution requires $\phi'(t_{c}) = \infty$, which indicates a singularity.

\section{Geodesics in homogeneous isotropic AdS cosmologies} \label{secgeos}

In this appendix we present the full computation of geodesics in crunching $AdS$ spacetimes.  As mentioned in \secref{georesults} the geodesic lengths only depend on the de Sitter invariant boundary separation. Thus, we break the calculation of geodesics into two sections: first, in \secref{secturnout} we compute the geodesics with sub-horizon boundary separation which never enter the  lightcone containing the crunching FLRW cosmology.  Then, in \secref{supHBS} we  compute antipodal geodesics which cover all super-horizon boundary separations.

\subsection{Sub-horizon boundary separation} \label{secturnout}
Without loss of generality we can take the boundary separation to be purely in the $\theta$ direction and set $\varphi = \dot{\varphi} = \ddot{\varphi} = 0$.  Then, the non-trivial geodesic equations outside the lightcone are:
\be \label{geos}
\begin{split}
	\ddot{\rho} &= a_{\rm out}(\rho)a_{\rm out}'(\rho)\left( \cosh^2{\tau}\dot{\theta}^2 - \dot{\tau}^2 \right) \\
	\ddot{\tau} &= -\cosh(\tau)\sinh(\tau)\dot{\theta}^2 - 2 {a_{\rm out}'(\rho) \over a_{\rm out}(\rho)} \dot{\rho}\dot{\tau} \\
	\ddot{\theta} &= -2\dot{\theta} \left( {a_{\rm out}'(\rho) \over a_{\rm out}(\rho)} \dot{\rho} + \tanh(\tau)\dot{\tau}\right)\,,
\end{split}
\ee
where dot denotes derivative with respect to an affine parameter. First, looking at the $\theta$ and $\tau$ equations and using the chain rule, $\ddot{\theta}=\ddot{\tau}\partial_\tau \theta+ \dot{\tau}^2 \partial_\tau^2\theta$, we can write the differential equation for $\theta(\tau)$:
\be 
{d^2\theta \over d\tau^2} = \cosh\tau\sinh\tau \({d\theta \over d\tau}\)^3 - 2 \tanh\tau {d\theta \over d\tau}\,.
\ee
This part of the geodesic motion is universal for any spacetime that can be written with de Sitter radial slices.  The solution is:
\be \label{thetaoftau}
\theta(\tau) = \arctan \left({\sinh(\tau) \over \sqrt{1+K \cosh^2(\tau)}}\right) - \arctan \left({\sinh(\tau_B) \over \sqrt{1+K \cosh^2(\tau_B)}}\right)\,,
\ee
where we have fixed one integration constant by making the arbitrary choice of initial conditions: the geodesic begins on the boundary at $\tau=\tau_B$ when $\theta=0$.

When $K<0$, geodesics turn back toward the boundary before reaching the lightcone.  In this case, the parametrization \eqref{thetaoftau} only covers the geodesic up to the turning point, because $\tau$ is not monotonic.  The turning point, $\tau_{\text{turn}}$, is given by the value of $\tau$ at which the parametrization fails:
\be \label{tauturn}
\tau_{\text{turn}} = \arccosh\sqrt{-{1\over K}}\,.
\ee
Plugging this back into \eqref{thetaoftau}, we see that the half-way point in the geodesic occurs at:
\be \label{thetaturn}
\theta_{\text{turn}} = {\pi \over 2} - \arctan \left( { \sinh\tau_B \over \sqrt{1+K\cosh^2\tau_B}}\right).
\ee

The radial equation contains all information about the potential through its dependence on $a_{\rm out}(\rho)$.  We write the differential equation for $\rho(\tau)$, and use \eqref{thetaoftau} to get:
\begin{equation} \label{e:e1}
0 = \frac{\D^2 \rho}{\D \tau^2} - 2 \left( \frac{\D \rho}{\D \tau} \right)^2 \frac{a_{\rm out}'(\rho)}{a_{\rm out}(\rho)} - \frac{\D \rho}{\D \tau} \frac{\tanh \tau}{1 + K \cosh^2 \tau} + \frac{a_{\rm out}(\rho) a_{\rm out}'(\rho) \cosh^2 \tau}{K^{-1} + \cosh^2 \tau}\,.
\end{equation}

To solve this equation we define a new function $\zeta = \zeta(\rho)$ and a new variable $\tau = \tau(\sigma)$ according to
\begin{align}
\zeta(\rho) & = \int_{\rho_0}^\rho \frac{\D r}{a_{\rm out}^2(r)}\,, \label{e:zrho} \\
\sigma(\tau) & = \arctanh \left( {\sqrt{K} \sinh(\tau) \over \sqrt{1+K\cosh^2(\tau)}} \right)\,. \label{sigoftau}
\end{align}
After these substitutions, the equation \eqref{e:e1} simplifies to
\begin{equation} \label{e:efin}
2 \frac{\D^2 \zeta}{\D \sigma^2} = \frac{\D}{\D \zeta} \frac{1}{a_{\rm out}^2(\zeta)}\,.
\end{equation}
The solution is:
\begin{equation} \label{e:efin2}
\sigma(\zeta) = \pm \int_{\z_0}^\z  \frac{ \D z}{\sqrt{E + a_{\rm out}^{-2}(z)}}\,,
\end{equation}
where $\z_0$ and $E$ are integration constants. By utilizing \eqref{e:zrho} we may express $\sigma$ in terms of the original radial variable, $\rho$, as
\begin{equation} \label{sigofrho}
\sigma(\rho) = \pm \int_{\rho_0}^{\rho}  \frac{\D r}{a(r) \sqrt{1 + E a_{\rm out}^2(r)}}\,,
\end{equation}
where $\rho_0 = \rho(\z_0)$ is the physical integration constant.

Due to the time reversal symmetry $\tau \mapsto -\tau$, we may assume $\tau_B > 0$, which fixes the sign in the above equation.  Since the integral in \eqref{sigofrho} converges at $r = \infty$, one can trade integration limit for an additive constant and rewrite
\begin{equation} \label{sigofrho2}
\sigma(\rho) = \sigma_{B} + \int_{\rho}^{\infty} \frac{\D r}{a_{\rm out}(r) \sqrt{1 + E a_{\rm out}^2(r)}}\,.
\end{equation}
The value of the integration constant is now fixed by enforcing
\be
\lim_{\rho \to \infty} \tau(\sigma(\rho)) = \tau_B\,.
\ee

As explained in \secref{geolentext}, it is sufficient to consider only the symmetric geodesics which reach the boundary at equal boundary time.  To select this class of geodesics we require that the position of the turning point in the $\rho$-direction occur at $\tau_{\rm turn}$ \eqref{tauturn}. The turning point in the $\rho$-direction is determined by the condition $\dot{\rho}|_{\rho_{\rm turn}} = 0$. Assuming that $\dot{\theta} \neq 0$, for non-antipodal geodesics, we find the condition:
\begin{align}
0 & = \left. \left( \frac{\D \rho}{\D \theta} \right)^2 \right|_{\rho_{\rm turn}} = \left. \left( \frac{\D \sigma}{\D \rho} \right)^{-2} \left( \frac{\D \sigma}{\D \tau} \right)^2 \left( \frac{\D \theta}{\D \tau} \right)^{-2} \right|_{\rho_{\rm turn}} \nn\\
& = a_{\rm out}^2(\rho_{\rm turn}) \left(E a_{\rm out}^2(\rho_{\rm turn}) + 1 \right) K \cosh^4 \tau\,.
\end{align}
Therefore, the turning point is given by:
\be \label{aturn}
a_{\rm out}(\rho_{\rm turn}) = \sqrt{-\frac{1}{E}}\,.
\ee
We see that if $E>0$ there is no turning point and the geodesics reaches the light cone.  For finite $E<0$ the geodesics turn outside the light cone.

Symmetric geodesics with equal boundary times satisfy the condition $\sigma(\tau_{\rm turn})=\sigma(\rho_{\rm turn})$.  Using \eqref{sigoftau} and \eqref{sigofrho2} and the turning points in the $\rho$ and $\tau$ directions, \eqref{aturn} and \eqref{tauturn}, we find the following relation between $K$, $\tau_B$ and $E$,
\begin{equation} \label{e:matching}
\frac{\I \pi}{2} - \arctanh \left( \frac{ \sqrt{K} \sinh \tau_B}{\sqrt{1 + K \cosh^2 \tau_B}} \right) = \int_{\rho_{\text{turn}}(E)}^{\infty} \frac{\D r}{a_{\text{out}}(r) \sqrt{1 + E a_{\text{out}}^2(r)}}\,.
\end{equation}

\subsection{Super-horizon boundary separation} \label{supHBS}
To find the regulated length as a function of boundary distance for super-horizon separations, we only need to consider antipodal geodesics.  In the limit $E\to\pm\infty$ the geodesic becomes the antipodal geodesic at the throat of de Sitter space with constant  $\tau =\tau_B=0$.  This geodesic never enters the lightcone and has a boundary separation equal to the de Sitter horizon, $D=\pi$ from \eqref{ZofD} .  As we increase $\tau_B$ from 0, antipodal geodesics cover all super-horizon boundary separations. 

From equation \eqref{thetaoftau} it follows that geodesics entering the light cone necessarily have  $K > 0$. For antipodal geodesics we can set $\theta =0$, and the geodesic equations simplify considerably. The requirement that $\theta(\tau) = 0$ for all $\tau$, combined with equation \eqref{thetaoftau}, implies $K = \infty$. Now the relation \eqref{sigoftau} simplifies to $\sigma = \tau$ and hence solutions to the geodesic equations outside the lightcone are given by:
\begin{equation} \label{e:taurho}
\tau(\rho) = \tau_B + \int_{\rho}^{\infty} \frac{\D x}{a_{\text{out}}(x) \sqrt{1 + E_{\text{out}} a_{\text{out}}^2(x)} }\,,
\end{equation}
and on the inside are given by:
\begin{equation} \label{e:chit}
\chi(t) = \int^{t_{\text{turn}}}_{t} \frac{\D x}{a_{\text{in}}(x) \sqrt{1 + E_{\text{in}} a_{\text{in}}^2(x)} }\,.
\end{equation}
Here, $E_{\text{out}}$ and $E_{\text{in}}$ are \textit{a prori} unrelated integration constants. We have fixed one integration constant in \eqref{e:taurho} by enforcing $\tau(\rho=\infty)=\tau_B$.  For the inner solution, \eqref{e:chit}, the limits of integration were chosen in such a way that $\chi(t_{\text{turn}}) = 0$; since we are considering antipodal geodesics, this selects only symmetric (equal boundary time) geodesics.  

The turning time, $t_{\rm turn}$ is given by:
\be
0 = \left.{d t \over d\chi}\right|_{t_{\rm turn}} = a_{\text{in}}(t_{\rm turn}) \sqrt{1 + E_{\text{in}} a_{\text{in}}^2(t_{\rm turn})}\,,
\ee
so $t_{\rm turn}$ is a solution of
\begin{equation} \label{e:pretbend}
a_{\rm in}(t_{\text{turn}}) =\sqrt{ - \frac{1}{E_{\text{in}}}}\,.
\end{equation}

At this point we must connect the solutions \eqref{e:taurho} and \eqref{e:chit} by matching across the lightcone.  To this end, we first use the analytic properties of the geodesic equations to generalize these expressions.  The geodesic equations, and their solutions on the inside and outside of the lightcone, are related by analytic continuation \cite{Dong:2011gx}. Since the geodesic equations are analytic, their solutions are analytic wherever defined, \textit{i.e.}, one can extend $a$ and $\phi$ to the complex  $t$ or $\rho$ plane. As in \eqref{e:aOutToIn}, we choose to use complex $t$, so that for real $t$ we have $a(t) = a_{\text{in}}(t)$ and $a(i t) = \I a_{\text{out}}(t)$.

First, we define the function
\begin{equation}
F_{E,z_0}(z) = \int_{z_0}^z \frac{\D x}{a(x) \sqrt{1 - E a^2(x)} }\,,
\end{equation}
where $z_0 \in \C$ is a fixed point. 
Equations \eqref{e:taurho} and \eqref{e:chit} can be rewritten as
\begin{equation} \label{geosofF}
\begin{split}
\tau(\rho) &= \tau_B + F_{E_{\rm out}, z_0}(\I \infty) - F_{E_{\rm out}, z_0}(\I \rho) \\
\chi(t) &= -F_{-E_{\rm in}, z_0}(t) + F_{-E_{\rm in}, z_0}(t_{\text{turn}}(E_{\text{in}}))\,.
\end{split}
\end{equation}
Next, to match the geodesic across the lightcone we can zoom in on the patch around $\rho, t=0$, where the space is locally flat, map our geodesics into Cartesian coordinates that cover both sides of the lightcone, and ensure that the geodesic is continuous and smooth.  The mapping to Cartesian coordiantes is done using the Rindler/Milne coordiatnes:
\be 
\begin{split}
X_{\text{out}} &= \rho \cosh \tau(\rho) \qquad X_{\text{in}} = t \sinh \chi(t) \\
T_{\text{out}} &= \rho \sinh \tau(\rho) \qquad T_{\text{in}} = t \cosh \chi(t)\,.
\end{split}
\ee
To expand \eqref{geosofF} around the lightcone we use the fact that for any non-singular $z_0$, the function $F$ is regular and its expansion around $z=0$ is 
\begin{equation}
F_{E,z_0}(z) = \log z + f_0 + \frac{1}{4} z^2 (E - 2 a_{0,3}) + \O(z^4)\,,
\end{equation}
where $a_{0,3}$ is the coefficient of $z^3$ in the series expansion\footnote{Here we have used the fact that $a(z)$ must be odd around the origin \cite{Dong:2011gx}.} of $a(z)$ and $f_0$ is a finite, $z_0$-dependent constant.  Then we find:
\be 
\begin{split}
X_{\rm out} &= {-\I A\over 2} + \rho^2\({-\I A \over 8}(-2a_{0,3}+E_{\rm out}) + {\I A^{-1}\over 2}\) + \O(\rho^4)\\
T_{\rm out} &= {-\I A \over 2} + \rho^2\({-\I A \over 8}(-2a_{0,3}+E_{\rm out}) - {\I A^{-1}\over 2} \)+ \O(\rho^4)\,,
\end{split}
\ee
and
\be 
\begin{split}
X_{\rm in} &= {1\over 2}B + t^2\({-B \over 8}(-2a_{0,3}-E_{\rm in}) - {B^{-1}\over 2}\) +\O(t^4)\\
T_{\rm in} &= {1\over 2}B + t^2\({-B \over 8}(-2a_{0,3}-E_{\rm in}) + {B^{-1}\over 2} \)+\O(t^4)\,,
\end{split}
\ee
where
\be \nn
A = \e^{F_{E_{\text{out}},z_0}(\I \infty)+\tau_B-f_0}, \,\,\,\text{and}\,\,\,\, B= \e^{F_{-E_{\text{in}},z_0}(t_{\text{turn}}) - f_0}\,.
\ee
Continuity requires that $X_{\rm in}(t=0)=X_{\rm out}(\rho=0)$ and implies that $-\I A = B$, or equivalently
\be
F_{E_{\text{out}},z_0}(\I \infty)+\tau_B - \I {\pi \over 2} = F_{-E_{\text{in}},z_0}(t_{\text{turn}})\,.
\ee
To enforce smoothness across the lightcone we additionally reiqure:
\be 
\begin{split}
\left.{d X_{\rm out} \over d T_{\rm out}}\right|_{\rho=0} &= \left.{d X_{\rm in} \over d T_{\rm in}}\right|_{t=0}\,, \\
\end{split}
\ee
which implies $ E_{\rm out} =- E_{\rm in}$.  By making the convenient choice $z_0 = \I \infty$, and defining $E=E_{\rm out}$, the geodesic solutions can be now written as
\be
\begin{split}
\tau(\rho) &= \tau_B - F(\I \rho) \\
\chi(t) &= F(t) - F(t_{\bend}(E)), \\
F(z) &= F_{E,\I \infty}(z) = \int_{\I \infty}^{z} \frac{\D x}{a(x) \sqrt{1 - E a^2(x)} }\,. \label{e:F}
\end{split}
\ee
The matching condition simplifies considerably to
\begin{equation} \label{e:tbend}
\tau_B = F \left( t_{\text{turn}} \right) + \frac{\I \pi}{2}, \,\,\,\,\,\text{and}\,\,\,\,\,\, a \left(t_{\text{turn}} \right) = \sqrt{\frac{1}{E}}\,.
\end{equation}

\section{Geodesic lengths} \label{seclengths}
In this section we  calculate the regulated lengths of the geodesic solutions found in appendix \ref{secgeos}. Our goal is to use the lengths in the relation \eqref{holodict} in order to gain some insight into the boundary QFT.  We start by simply writing the lengths of geodesics for both the case of sub- and super-horizon separation treated in \secref{secgeos}.  For geodesics that never enter the light cone containing the crunching FLRW cosmology we have:
\begin{equation} \label{lenout}
\L = 2\int_{\rho_{\rm turn}}^{\rho_{\rm cut}}{\sqrt{E_{\rm out}}a_{\rm out}(\rho)d\rho \over \sqrt{1+E_{\rm out}a_{\rm out}^2(\rho)}}\,,
\end{equation}
and for antipodal geodesics we have:
\begin{equation}
\L_{\rm reg} = 2\(\int_0^{\rho_{\rm cut}}{\sqrt{E_{\rm out}}a_{\rm out}(\rho)d\rho \over \sqrt{1+E_{\rm out}a_{\rm out}^2(\rho)}} + \int_0^{t_{\rm turn}}{\sqrt{-E_{\rm in}}a_{\rm in}(t)dt \over \sqrt{1+E_{\rm in}a_{\rm in}^2(t)}}\),
\end{equation}
where $\rho_{\rm cut}$ is a near-boundary cut-off.

These two cases can be unified using the information \eqref{e:aOutToIn}, and the conventions defined in \secref{supHBS}: $E=E_{\rm out}=-E_{\rm in}$, $a(t)=a_{\rm in}(t)$, and $-a(t) = \I a_{\rm out}(\I t)$.  Upon substitution we find:
\be \label{lenregfin}
\L = 2\int_{t_{\rm cut}}^{t_{\rm turn}} {\sqrt{E}a(t) dt \over \sqrt{1-Ea^2(t)}},
\ee
where the $t$ cut-off is related to the one above by $t_{\rm cut}= i \rho_{\rm cut}$.  Here, as always, $t_{\rm turn}$ is given by:
\begin{equation} \label{e:bend}
a(t_{\rm turn}) = \sqrt{\frac{1}{E}}.
\end{equation}
For $E > 1$ the turning point occurs inside of the lightcone and $t_{\rm turn}$ matches \eqref{e:pretbend}. For $E < 0$ the solution to \eqref{e:bend} becomes purely imaginary and in such a case the turning occurs outside of the lightcone at positive $\rho_{\rm turn} = \pm \I t_{\rm turn} $, in agreement with \eqref{aturn}.

\subsection{The de Sitter invariant distance} \label{app:dSdist}
The reader may be concerned with the fact that we have thus far only obtained expressions for the length in terms of an integration constant, $E$, whereas we know that the length is a function of the de Sitter invariant distance between boundary points.

For 3-dimensional de Sitter  space  in the following coordinates,
\begin{equation}
\D s^2 = - \D \tau^2 + \cosh^2 \tau ( \D \theta^2 + \sin^2 \theta \D \varphi^2 )\,,
\end{equation}
the invariant $Z$ \eqref{Zdef} is:
\begin{equation} \label{e:Z}
Z = -\sinh(\tau_1)\sinh(\tau_2) + \cosh(\tau_1)\cosh(\tau_2)\cos(\Delta \theta)
\end{equation}
where we assume $\Delta \varphi = 0$, and have set the de Sitter radius to 1.

A relation between the integration constant, $E$, and the de Sitter distance $D$ or $Z$ (related by \eqref{ZofD}) can be obtained. Substituting the angular distance on the boundary, given by equation \eqref{thetaturn}, into \eqref{e:Z} and choosing equal boundary times, yields
\begin{equation} \label{e:ZoftauB}
Z = - \frac{1 + K \cosh(2 \tau_B)}{1 + K}\,.
\end{equation}
Using the relation between integration constants, \eqref{e:matching}, obtained by selecting only symmetric geodesics, we substitute $K(\tau_b, E)$ into \eqref{e:ZoftauB}.  The result is especially simple in terms of the geodesic distance $D$:
\begin{equation} \label{e:DE}
\frac{\I D}{2} = \int_{\rho_{\text{turn}}(E)}^{\infty} \frac{\D r}{a_{\text{out}}(r) \sqrt{1 + E a_{\text{out}}^2(r)}}\,.
\end{equation}
The dependence on $\tau_B$ cancels, and we find that boundary separation is a function of $E$ alone for symmetric geodesics.

In case of antipodal geodesics the formulae simplify even more. With $K = \infty$ equation \eqref{e:ZoftauB} reads
\begin{equation}
Z = - \cosh(2 \tau_B)\,.
\end{equation}
Together with equation \eqref{e:tbend} and \eqref{ZofD} we immediately obtain
\begin{equation} \label{e:DE2}
\frac{D}{2} = F(t_{\text{turn}}(E)) + \frac{1}{2}( 1 + \I ) \pi\,.
\end{equation}

Both equations \eqref{e:DE} and \eqref{e:DE2} can be expressed in terms of complex function $F$ defined in \eqref{e:F} and complexified scale factor $a(z) = a_{\text{in}}(z)$. In terms of $Z$ a single equation can be written down,
\begin{equation} \label{Zoftturn}
Z = \cosh \left[ 2 F(t_{\bend}) \right].
\end{equation}
This equation can be now used to relate the boundary separation $D$ or $Z$ to the integration constant $E$ via:
\begin{equation}
a^2(t_{\turn}) = \frac{1}{E}\,.
\end{equation}
In this sense one could treat $t_{\rm turn}$ as an integration constant and eliminate $E$ entirely.

\subsection{Limiting behavior}

While the precise form of the dependence of the geodesic length, $\mathcal{L}$, on the boundary separation, $D$ or $Z$,  is model-dependent, the behavior is universal in certain limits. In particular, for very small boundary separations $D \rightarrow 0$ or $Z \to 1^-$, geodesics only probe the near-boundary geometry, $\rho_{\rm turn} \gg 1$, and hence their length should approach the length of geodesics in pure $AdS$. 

As explained in \secref{geolentext}, the divergent geodesic lengths are regulated  with respect to a cut-off in the conformal factor to the boundary metric, $a_{\rm out}(\rho)$.  As the turning point, $\rho_{\rm turn}$, approaches infinity, the value of the scale factor at the turning point, $a_{\rm turn}=a_{\rm out}(\rho_{\rm turn})$ diverges.  In the near boundary limit, we expand \eqref{e:DE} in $a_{\rm turn}$:
\be \label{ZDofturn}
\begin{split}
 D  &= \sqrt{2(1-Z)} +\O(1-Z)^{3/2} \\
& ={2\over a_{\rm turn}} \(1- {2+\alpha^2 \over 6~ a_{\rm turn}^2}\) + \O(a_{\rm turn}^{-5})\,.
\end{split}
\ee
This is accomplished via a change of variables which puts \eqref{e:DE} in terms of the scale factor, using the equations of motion $d a_{\rm out}/d\rho = \sqrt{a_{\rm out}^2(\phi'^2/6-V/3)+1}$, and the asymptotic expansion of $\phi$ \eqref{aysphi}. 

Similarly, we can write the length \eqref{lenout} in terms of the scale factor; in the near boundary expansion the relation between the length of geodesics and the scale factor at the turning point is
\begin{equation} \label{lenofturn}
\mathcal{L} = 2(\log(a_{\rm cut})-\log(a_{\rm turn}))+\log4 - \frac{1+\alpha^2/2}{a_{\rm turn}^2} + \mathcal{O}(a_{\turn}^{-3})\,.
\end{equation}
Solving \eqref{ZDofturn} for $a_{\rm turn}$ and inserting it in \eqref{lenofturn} one obtains
\begin{align}
\mathcal{L} &=2\log(a_{\rm cut})+ \log\( D^2\) + \frac{2+\alpha^2}{24}D^2 + \mathcal{O}(D^4) \\
&= 2\log(a_{\rm cut})+ \log \(2(1-Z) \) -{\alpha^2\over 12}(1-Z)+\mathcal{O}(1-Z)^2\,.\label{lregZsm}
\end{align}
From this expression we arrive at the subtraction scheme \eqref{renscheme}:
\be \label{renapp}
\begin{split}
\L_{\rm reg} &= \lim_{a_{\rm cut}\to \infty} \( \L_{\rm reg} -2\log(a_{\rm cut}) \) \\
&= \lim_{\rho_{\rm cut}\to \infty} \(\L_{\rm reg} -2\rho_{\rm cut} \) -\log(a_1^2) \, .\\ 
\end{split}
\ee	

Another interesting limit occurs for large boundary separations: $D \rightarrow \infty$, $Z\to -\infty$. In this limit geodesics turn on the inside of the lightcone and the turning point $t_{\turn}$ approaches the time $t_{\rm max}$ where the scale factor reaches its maximum value, $a_{\rm max} = a(t_{\rm max})$. At this point $a'(t_{\rm max}) = 0$ and  $a''_{\rm max} \equiv a''(t_{\rm max}) < 0$.  Expanding the integral in \eqref{e:DE2} around $t = t_{\rm max}$, one can find the leading term in a relation between $D$ and $t_{\turn}$ at large $D$,
\begin{equation} \label{DlgD}
D = \log\( -2Z \) + \mathcal{O}(Z^0) = \frac{-1}{\sqrt{-a_{\rm max} a''_{\rm max}}} \log (t_{\rm max} - t_{\turn})^2 + \mathcal{O} \left((t_{\rm max} - t_{\turn})^0 \right).
\end{equation}
Similarly for the length \eqref{lenregfin} becomes:
\begin{equation}
\mathcal{L_{\rm reg}} = - \sqrt{\frac{a_{\rm max}}{-a''_{\rm max}}} \log (t_{\rm max} - t_{\turn})^2 + \mathcal{O} \left((t_{\rm max} - t_{\turn})^0 \right),
\end{equation}
which leads to
\begin{align}
\mathcal{L_{\rm reg}} &= a_{\rm max} D + \mathcal{O}(D^0) \\
&= a_{\rm max} \log \(-2Z \)+ \mathcal{O}(Z^0) \label{lregZlg}.
\end{align}

In this case, finite terms depend on the specific dynamics and cannot be extracted exactly.  However, we are able to estimate the constant contribution to \eqref{lregZlg}.  First, we note that there is a universal behavior of the length for the antipodal geodesic with horizon separation: as noted in \secref{supHBS} the antipodal geodesic at $\tau_B=0$ corresponds to the limit $E\to -\infty$.  In this limit the length integral \eqref{lenout} simplifies to:
\be
\L = 2\int_{\rho_{\rm turn}}^{\rho_{\rm cut}} d\rho = 2\rho_{\rm cut}\,,
\ee
because $\rho_{\rm turn}(E\to \infty)=0$.  Thus, after the subtraction \eqref{renapp}, we see that the regulated length at the horizon is given by $-\log(a_1^2)$.  

We then approximate the constant term in \eqref{lregZlg} by approximating the super-horizon dependence to be purely linear in $D$ with a slope given by $a_{\rm max}$:
\be \label{Llinear}
\L_{\rm reg} \approx (D -\pi) a_{\rm max}  - \log(a_1^2)\,.
\ee
Numerics show this approximation is accurate at the sub-percent level for $\phi_0 \gtrsim 10$.

\bibliography{refs}

\end{document}